\documentclass[lettersize,journal]{IEEEtran}
\usepackage{amsmath,amsfonts}
\usepackage{algorithmic}
\usepackage{algorithm}
\usepackage{array}
\usepackage[caption=false,font=normalsize,labelfont=sf,textfont=sf]{subfig}
\usepackage{textcomp}
\usepackage{stfloats}
\usepackage{url}
\usepackage{verbatim}
\usepackage{graphicx}
\usepackage{cite}
\begin{document}

\title{Self-supervised Physics-based Denoising for Computed Tomography}

\author{Elvira~Zainulina, 
        Alexey~Chernyavskiy, 
        Dmitry~V.~Dylov
\thanks{Elvira~Zainulina and Alexey~Chernyavskiy are with the Philips Innovation Labs, Moscow, Russia.}
\thanks{Elvira~Zainulina and Dmitry V. Dylov are with the Skolkovo Institute of Science and Technology, Bolshoy blvd., 30/1, Moscow 121205, Russia}
\thanks{Corresponding author: d.dylov@skoltech.ru}}

\markboth{Zainulina \MakeLowercase{\textit{et al.}}: Self-supervised Physics-based Denoising for Computed Tomography}%
{Zainulina \MakeLowercase{\textit{et al.}}: Self-supervised Physics-based Denoising for Computed Tomography}


\maketitle

\begin{abstract}
Computed Tomography (CT) imposes risk on the patients due to its inherent X-ray radiation, stimulating the development of low-dose CT (LDCT) imaging methods.
Lowering the radiation dose reduces the health risks but leads to noisier measurements, which decreases the tissue contrast and causes artifacts in CT images. 
Ultimately, these issues could affect the perception of medical personnel and could cause misdiagnosis.
Modern deep learning noise suppression methods alleviate the challenge but require low-noise~-- high-noise CT image pairs for training, rarely collected in regular clinical workflows.

In this work, we introduce a new self-supervised approach for CT denoising Noise2NoiseTD-ANM that can be trained without the high-dose CT projection ground truth images. 
Unlike previously proposed self-supervised techniques, the introduced method exploits the connections between the adjacent projections and the actual model of CT noise distribution. Such a combination allows for interpretable no-reference denoising using nothing but the original noisy LDCT projections.
Our experiments with LDCT data demonstrate that the proposed method reaches the level of the fully supervised models, sometimes superseding them, easily generalizes to various noise levels, and outperforms state-of-the-art self-supervised denoising algorithms.
\end{abstract}


\section{Introduction}
The concerns about the potential health risks caused by the X-ray radiation of computed tomography (CT) have led to the concept of ALARA (As Low As Reasonably Achievable)~\cite{international19921990},~\cite{slovis2003children},~\cite{brenner2007hall},~\cite{SARMA2012750}. 
The concept aims to regulate the delivery of excessive X-ray radiation to the patients. 
Although low-dose CT (LDCT) imaging decreases the risks, it yields images of deteriorated quality because of the increased noise background and the pertinent appearance of artifacts that could affect the diagnostic decisions~\cite{yu2009radiation}.

For mitigating the excessive noise in LDCT, different methods were proposed. One of the approaches is to apply iterative reconstruction techniques. These methods aim to improve the quality of CT slices through optimization of an objective function that incorporates a system model, a statistical noise model, and prior information in the image domain. Although these iterative techniques gave an improvement in the quality of the reconstructed CT slices, they are not efficient because they require multiple forward and back-projection operations.

\begin{figure}[!t]
\centering
\includegraphics[width=2.5in]{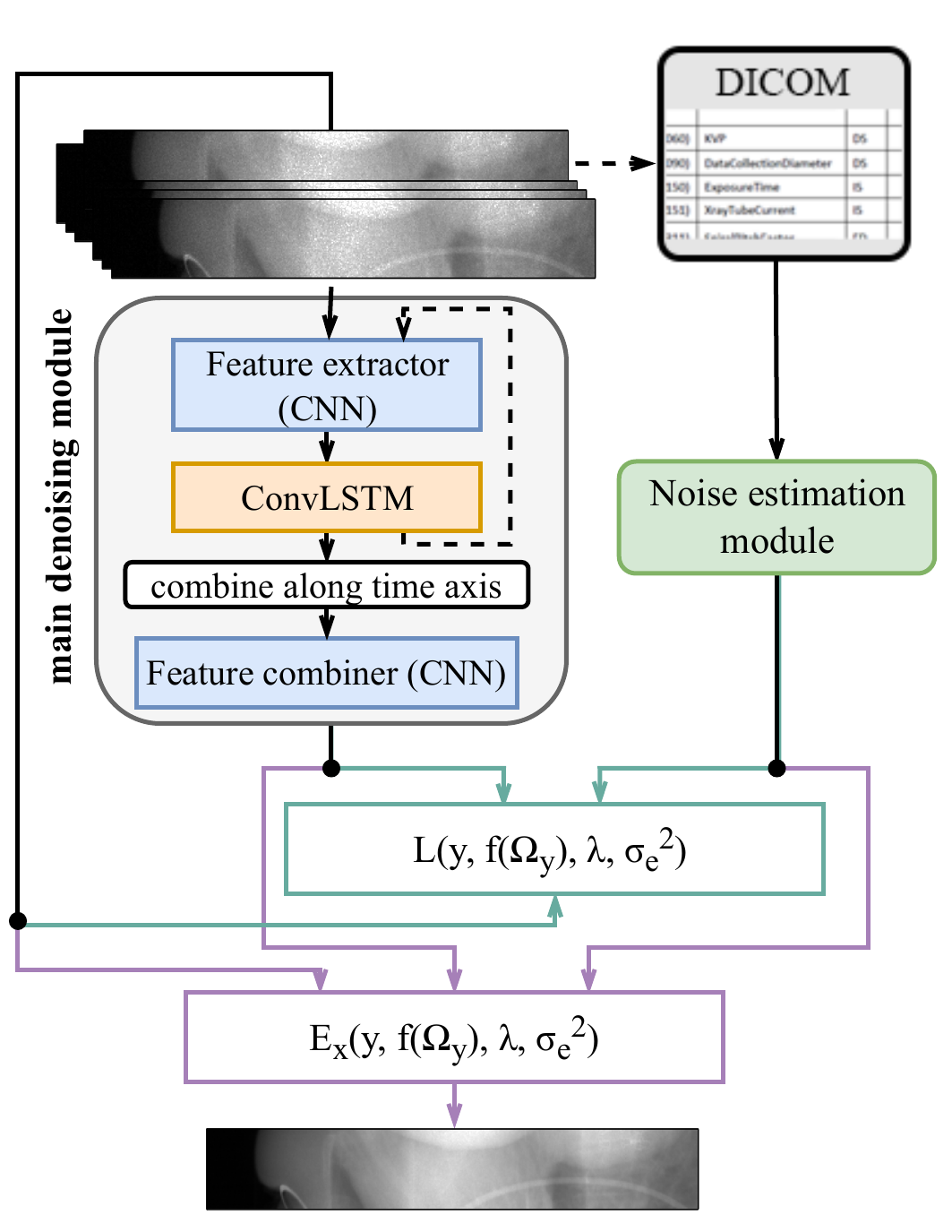}
\caption{The scheme of the proposed self-supervised Noise2NoiseTD-ANM approach. $y$ denotes noisy pixels, $\Omega_y$ is neighborhood of noisy pixels, $\lambda$ and $\sigma_e^2$ are estimated noise parameters. Green arrows correspond to the training process and purple arrows correspond to inference.}
\label{fig:approach}
\end{figure}

The other approach is to employ image processing: either traditional or deep-learning-based. Because these methods were adapted from denoising images in the natural domain, they are not overly dependent on the CT scanner and its acquisition geometry. Compared to iterative reconstruction techniques, they can separately denoise raw CT projections or already reconstructed CT slices. These methods can serve as an auxiliary tool for noise reduction before or after reconstruction. Thus, they, especially deep-learning-based methods, are of high research interest.
The traditional methods are not very adaptive to image content and often contain many parameters which require careful case-by-case tuning. Besides, they provide inferior quality than deep-learning-based methods. 
Specifically, supervised deep learning methods (Noise2Clean) that assume model training on the big data set of paired noisy-clean images show the best performance~\cite{3DCT},~\cite{RED},~\cite{ResCT}. However, their requirement for the paired noisy-clean images may not always be accomplished.  
In a clinical setting, for example, in a radiology suite, the acquisition of such paired data can take at least twice longer than the regular CT exam and would increase exposure to the X-ray dose. Furthermore, obtained images may not ideally coincide due to the patient breathing and movements.
Generating the noisy data artificially is an option to alleviate the lack of the image pairs, but the addition of the noise to the images, especially the medical ones, can bias the predictions of a convolutional neural network (CNN)~\cite{green2018learning} unless a very accurate noise simulation is used, which requires deep insights into the actual CT system,~\cite{zabic2013low} and is not always possible.

Because of the problems with the data availability, different unsupervised and self-supervised methods for image denoising were proposed. However, unsupervised methods, which are mostly based on generative adversarial networks~\cite{park2019unpaired},~\cite{wolterink2017generative},~\cite{kang2019cycle}, sill require a set of clean images. Besides, this set, as well as the set of noisy images, should be well representative of the diversity of cases found in reality. Thus, these methods imply the need to irradiate some patients with rather high doses of CT X-rays. 
Self-supervised methods, which do not require clean images at all, appear to be more suitable for denoising medical images than supervised and unsupervised methods. 

Some of the previously proposed self-supervised CT denoising methods~\cite{n2n_ct2020},~\cite{wu2019consensus},~\cite{hendriksen2020noise2inverse},\cite{yuan2020half2half} require additional data generation that complicates the denoising process. The other methods~\cite{xu2021deformed2self},~\cite{choi2021self},~\cite{unal2021self} do not have this requirement. However, the previous self-supervised CT denoising methods do not get advantage from all the available information, such as CT noise properties, and some of them do not consider similarity of adjacent projections. A more detailed description of related work on self-supervised denoising will be given in the section~\ref{sec:rel_work}.
In this paper, we propose a new self-supervised approach for denoising that takes into account the mentioned properties.

Unlike most previously proposed methods that denoise already reconstructed CT slices, we propose to work in the projection domain. Although radiologists work with CT slices, image processing methods can be directly applied to projections. Working with CT projections may be more advantageous than processing CT slices. Firstly, CT reconstruction is an ill-posed inverse problem. There are a lot of different reconstruction methods, which have many parameters influencing quality of reconstructed CT slices. A method developed for a one set of parameters can be unsuitable for other sets leading to the necessity of stack of methods designed for each set. For the projection data, there is no such a problem. Secondly, after the reconstruction CT slices may contain artefacts, e.g., streaks, caused by the noise in projections, which are more difficult to be removed than the initial noise. 
Thirdly, in the case of deep learning methods, they require many data for training. Here is one more benefit to working in the projection domain, as the projection data of one patient contains more images than the reconstructed image data of this patient. Because the medical images are difficult to be obtained, it is a crucial property. 
Finally, CT projections share almost the same content, and the noise is spatially uncorrelated
on projections that can be useful for denoising, especially in the absence of clean images. One more point about noise is that its pixel-wise independent distribution on projections is one of the main assumptions for the possibility to apply self-supervised denoising methods. In the image domain, this assumption is violated because of the reconstruction process. 

In this paper, we propose the self-supervised Noise2NoiseTD-ANM (ANM stands for ``adaptable noise model'') approach for denoising low-dose CT projections that is an extension of our earlier proposed method in~\cite{zainulina2021n2ntd}. Like our previous method, it uses information contained in sequences of images, whereas most other self-supervised denoising methods denoise each image separately, which can give sub-optimal quality of denoised images. In addition, the developed approach takes into account the theoretical distribution of the CT noise. This makes denoising models better generalize to different noise levels, even relatively high ones.
The scheme of the proposed approach is illustrated in Fig.~\ref{fig:approach}.


\subsection{Contributions}
The contribution of this paper is in the following:
\begin{enumerate}
    \item Novelty. The approach incorporates actual CT noise model. The noise model improves the capability of our denoising framework to generalize to previously unseen noise levels.
    \item Flexibility. Our approach simplifies model adaptation to different CT scan settings because it is designed to pre-estimate parameters of the noise model and to optimize the denoising and the noise models separately. 
    \item Validation. We performed the comparison of the proposed approach, the Noise2Clean, the Half2Half technique~\cite{yuan2020half2half}, and one of the best self-supervised methods introduced in the paper ``High-Quality Self-Supervised Deep Image Denoising'' \cite{laine2019hqss}, adapted to CT projection denoising. We used both simulated data (three different noise levels) and a dataset with the real noise. Our method outperformed state-of-the-art algorithms.
\end{enumerate}

\section{Related work}
\label{sec:rel_work}
The first attempt to train a neural network using only noisy images was made by J. Lehtinen et al. Their work~\cite{Noise2Noise} gave rise to self-supervised denoising methods. Their method consists in training a neural network to predict one noisy image from another. The approach is based on the properties of the loss functions, which allow the model to converge in the limit to the same state as in the case of supervised training. For the applicability of the approach, the noise have to be zero-mean and noise components of different pixels have to be independent to each other.
For medical data, this approach does not seem appropriate because of the requirement of paired noisy images. However, different methods were introduced that extend the Noise2Noise approach and adopt it to denoise CT images. Some of them give solutions on how to generate an acceptable dataset for model training. 

In~\cite{n2n_ct2020} the Noise2Noise approach was applied for denoising X-ray projections and CT images and compared to the Noise2Clean. The Noise2Noise approach has shown rather good results compared to the Noise2Clean. However, they used an artificially generated dataset for the experiments. 

In the works~\cite{wu2019consensus} and~\cite{hendriksen2020noise2inverse}  the ``data splitting'' methods were proposed for generation of a dataset for the Noise2Noise approach. These methods consist in reconstruction of paired CT slices from non-intersecting sets of CT projections. Although these methods showed rather good results, they may introduce reconstruction artifacts that can violate the noise assumptions. The other possible negative outcome of the ``data splitting'' is the drop in resolution of reconstructed images. These methods allow denoising only already reconstructed slices. 

\cite{yuan2020half2half} introduced a Half2Half method consisting in the obtainment of two half-dose projections from the given projection using the theoretical distribution of count-domain data for Noise2Noise training. The method gave acceptable results in the experiments carried out by the authors and provided a reasonable approach for the dataset generation. 
However, the Half2Half approach needs additional knowledge about acquisition parameters for noise simulation that is not always possible to obtain. 

Thus, the Noise2Noise-based methods have the main bottleneck consisting in accurate data generation satisfying all assumptions of the Noise2Noise. This bottleneck makes the models less flexible. 

The drawback with the dataset generation is overcome by self-supervised methods based on the usage of blind spots. These methods assume model training using only original noisy images as input and target. Nevertheless, the application of the blind-spot-methods for denoising of low-dose CT projections is little researched. Some of the methods use pixel masking during training~\cite{krull2019noise2void},~\cite{xie2020noise2same}, which makes them computationally inefficient, and other use special neural network architectures~\cite{laine2019hqss},~\cite{batson2019noise2self},~\cite{lee2020noise2kernel} that restricts their flexibility. The main weakness of most blind-spot-based methods is that they exclude information about the noisy pixel to be denoised that makes them prone to loss of small details and blurring and cause worse performance compared to the Noise2Clean and Noise2Noise approaches. One of the most efficient solutions for this problem 
was introduced in~\cite{laine2019hqss}, where the authors proposed to include information about the excluded noisy pixel during inference by Bayes rule.
This approach was adopted by us for denoising of low-dose CT projections in~\cite{zainulina2021n2ntd}. Also, in~\cite{zainulina2021n2ntd} we proposed a new method Noise2NoiseTD, which uses information from adjacent projections that can help prevent the model from over-smoothing and losing edges.

Recent works~\cite{xu2021deformed2self},~\cite{choi2021self},~\cite{unal2021self} also attempt to perform self-supervised denoising of CT data, however they do not take the true noise model into account.

In the paper, we propose the extension of our previous approach. 
We will compare it to the approach from~\cite{laine2019hqss} adopted for denoising of CT projections, which will be referred to as Noise2Void-4R ($4$R stands for $4$ rotations), and to the Half2Half approach~\cite{yuan2020half2half}.

\section{Methods}
In this section, we present the description of our approach. Firstly, we give an overview of the CT noise properties. Then, we describe how we use the adjacency in the approach and how we incorporate it in a neural network. After that, we describe the train-inference schemes of the model depending on the CT noise distribution. We introduce a noise model for self-supervised training that incorporates the CT noise properties. The description of the usage of the adjacency and the noise model completes the description of the Noise2NoiseTD-ANM approach. Additionally, in the section, we present the neural network architecture used as the backbone for the Noise2NoiseTD-ANM, Noise2Void-4R, Half2Half, and Noise2Clean approaches for comparison. There will be given details about modifications of the neural network corresponding to the applied approaches.

\subsection{CT noise properties}
\label{sev:CT_noise}
CT projections are the line integrals of the linear attenuation coefficients of the body. They are obtained after normalization applied to the data measured by detectors.  

According to~\cite{buzug2011computed}, the measured data can be described by Lambert-Beer's law:
\begin{equation}
    I = I_0\exp(-p),
\end{equation}
where $I$ is the number of detected photons, $I_0$ is the incident number of photons, $p$ is the line integral of linear attenuation coefficients, i.e., the projection.
Let $\exp(-p)$ be the transmission data $T$. Then, $I=I_0T$.

In practice, the actual value of $I$ is not available: it is corrupted by noise. The detected number of photons $I$ is a random variable that can be described by Poisson distribution ($\mathcal{P}$)~\cite{buzug2011computed}. Besides, the electronic noise inherent in the CT scanner contributes to the overall noise level. The electronic noise can be modeled by Gaussian distribution with parameters $\mu_e$ and $\sigma_e^2$, the mean and variance of the electronic noise, $\mathcal{N}\left(\mu_e, \sigma_e^2\right)$~\cite{la2006penalized}. In CT systems, $\mu_e$ is usually calibrated to be $0$. Thus, the detected number of photons obey the following mixed Poisson-Gaussian distribution:
\begin{equation}
    I = \mathcal{P}\left(I_0T_{hd}\right) + \mathcal{N}\left(0, \sigma_e^2\right).
\end{equation}
$T_{hd}=\exp(-p_{hd})$ and $p_{hd}$ are the transmission and projection data, respectively, which do not contain noise ($hd$ corresponds to high-dose data). 

Then, for the noisy projection $\hat{p}$, the transmission data $\hat{T}$ can be modeled by the distribution:
\begin{equation}
    \hat{T} = \exp(-\hat{p}) = \frac{I}{I_0} = \frac{1}{I_0}\left(\mathcal{P}\left(I_0T_{hd}\right) + \mathcal{N}\left(0, \sigma_e^2\right)\right).
\end{equation}

Approximating Poisson noise as signal-dependent Gaussian noise and defining high-dose transmission data as $x=T_{hd}=\exp(-p_{hd})$, we can express the transmission data for the noisy projection using Gaussian distribution:
\begin{equation} \label{eq:distr}
    \hat{T} = \exp(-\hat{p}) = \mathcal{N}\left(\mu_x, \sigma_x^2+\frac{\mu_x}{I_0}+\frac{\sigma_e^2}{I_0^2}\right).
\end{equation}

Whereas $\mu_x$ and $\sigma_x^2$ characterize clean data, $I_0$ and $\sigma_e^2$ are the noise parameters depending on the properties of a CT scanner. The commonly employed in CT scanners bowtie filtering modulates an X-ray beam as a function of the angle to balance the photon flux on a detector array~\cite{liu2014dynamic}. It reduces the radiation at the periphery of the field of view resulting in a bell shape distribution of incident flux levels. Besides, recent CT scanners use a dose modulation technique,
which regulates the tube current for each projection angle depending on the properties of organs to not expose to excessive radiation. Because the flux level is proportional to the tube current, dose modulation influences its distribution. Thus, the incident flux level and, hence, the noise variance depends on the position of the detector column and the tube current.
As for the variance of electronic noise $\sigma_e^2$, it is the characteristic of the detector and does not depend on the X-ray beam. 

\subsection{Using adjacent projections in a time-distributed denoising model}
As in our previous work~\cite{zainulina2021n2ntd}, we propose to restore the projection depending on its noisy adjacent projections. However, the previously proposed model was not completely blind causing its over-fitting to the noise at some moment during training and requiring early stopping. In this work we exclude the projection to be denoised from the consideration of the denoising model.
The problem can be considered as the prediction of the middle frame in the sequence depending on its past and future frames. Because the neighboring projections share almost the same content, while the noise is independently distributed on each projection, the prediction will be mostly dependent on the structural features of the projections, excluding the noise features. It will allow recovering noise-free middle frame. Thus, we restore a denoised version of a projection $p_i$ from projections $p_{i\pm 1},\dots,p_{i\pm k}$. The choice of the number of the adjacent frames $k$ depends on how much the content on two consequent frames differs and the computational and the memory capacity of the device. In this study $k=3$ is chosen.

In order to predict the frame from the sequence of past and future frames, we propose using the convolutional long short-term memory (ConvLSTM) units because LSTM was shown to be effective in the time-series tasks. ConvLSTM units included in the model are based on the ConvLSTM layers introduced in~\cite{shi2015convlstm}. Compared to~\cite{shi2015convlstm}, we do not use the past cell status for gates calculation, reducing the number of model parameters without really affecting prediction quality. Also, according to~\cite{bias-free}, all the convolutions are made bias-free. In this study, the ConvLSTM unit consisted of one ConvLSTM cell.

This unit processes features extracted by some CNN for each frame independently.  
It makes a prediction in one direction, i.e. the middle projection $p_i$ is predicted from the combination of the features extracted from the sequence $p_{i-k},\dots,p_{i-1}$ and from the sequence $p_{i+1},\dots,p_{i+k}$. The features extracted from both sequences are combined using attention~\cite{SE} along the time axis and concatenation along the channel axis. Then, these features are combined by a fusing CNN to obtain the denoised middle projection.

\subsection{Train-inference scheme}

As in~\cite{zainulina2021n2ntd}, we use the train-inference scheme proposed in~\cite{laine2019hqss} that assumes model training and inference depending on the Gaussian approximation of the data distribution.
According to the description of the CT data distribution, the noisy and clean transmission data can be modeled by Gaussian distribution with rather explainable parameters. Therefore, the training of the models for the self-supervised approaches is carried out on the transmission data.

Let  $y=\left(y_1,\dots,y_N\right)$ denote the pixels of noisy transmission data. Each noisy pixel has a neighborhood $\Omega_{y_i}$ that consists of noisy pixels surrounding the pixel $y_i$ on its adjacent projections. Let $\Omega_y=\left(\Omega_{y_1},\dots,\Omega_{y_N}\right)$ be the neighborhoods of the noisy pixels.
We want to restore the pixels of clean data $x=\left(x_1,\dots,x_n\right)$.
The parameters of the Gaussian distribution of clean data $p(x|\Omega_y)$ are predicted by neural networks. Because only the noisy data $\mathcal{D}=\left\{y_i, \Omega_{y_i}\right\}_{i=1}^N$ is available, then the networks are trained using a loss function that maximizes the log-likelihood of the distribution of the noisy data $p(y|\Omega_y)$:
\begin{equation}
    \mathcal{L}=-\sum_{i=1}^{n}\log{p(y_i|\Omega_{y_i})}.
\end{equation}
After substituting the distribution from~\ref{eq:distr}, we get the following loss function:
\begin{equation} 
\label{eq:loss}
\begin{split}
    \mathcal{L}=\sum_{i=1}^{n}\left(\frac{(y_i-\mu_{x_i})^2}{2\sigma_{y_i}^2} + \frac{1}{2}\log\sigma_{y_i}^2\right),\\
    \sigma_{y_i}^2 = \sigma_{x_i}^2+\sigma_{n_i}^2,\; 
    \sigma_{n_i}^2 = \frac{\mu_{x_i}}{\lambda}+\frac{\sigma_e^2}{\lambda^2}.
\end{split}
\end{equation}
The $\lambda$ denotes the parameter approximating the actual incident flux level $I_0$. The $\sigma_e^2$ denotes approximation of the electronic noise variance.

With this loss function, the model is trained to predict estimations of the parameters of the clean data distribution $\mu_x$ and $\sigma_x^2$. Then we get the prediction of the clean data incorporating the information about the noisy pixels $y$ itself by Bayes rule:
\begin{equation}
    p(x|y, \Omega_y) \sim p(y|x)p(x|\Omega_y).
\end{equation}
Given $p(y|x)=\mathcal{N}_y(x, \sigma_n^2)=\mathcal{N}_x(y, \sigma_n^2)$ (the equation used the symmetry of Gaussian distribution to swap $x$ and $y$) and $p(x|\Omega_y)=\mathcal{N}(\mu_x,\sigma_x^2)$ are normal distributions, the distribution $p(x|y, \Omega_y)$ is a scaled normal with the mean~\cite{bromiley2003products}:
\begin{equation} \label{eq:prediction}
    \mathbb{E}_x[p(x|y, \Omega_y)] = \frac{y\sigma_x^2+\mu_x\sigma_n^2}{\sigma_x^2+\sigma_n^2}.
\end{equation}
This mean coincides with the posterior mean estimation of the clean data that is used for prediction.

\subsection{Noise model}
 We propose to make the neural network representing the noise model with the parameters $\lambda$ and $\sigma_e^2$ that approximate the incident flux level and the electronic noise variance, respectively. The parameters of the noise model can be pre-estimated independently and optimized together with the main denoising model. The noise model parameters can be pre-estimated depending on the CT properties described earlier.
 
As for the parameter $\lambda$, if the bowtie filtering was applied, the $\lambda$ parameter should differ across detector columns, i.e., it should depend on the column position of the pixel to be denoised. Otherwise, some parts of the projection would be over-smoothed, and others would stay noisy because of the under- or overestimation of the Poisson parameter of the noise. 
Because the modern CT scanners use dose modulation, the $\lambda$ parameter should also depend on the used tube current. Thus, $\lambda=\lambda(i, mA)$, where $i$ is a detector column number and $mA$ is a tube current.
As for the variance of electronic noise $\sigma_e^2$, we assume that $\sigma_e^2$ is the same for all pixels. It probably differs between the detector cells, but we assume that this difference is relatively small. The possible error is negligible compared to the cost caused by the increase in the number of parameters by the number of detector cells.

\begin{figure}[!t]
\centering
\includegraphics[width=2.5in]{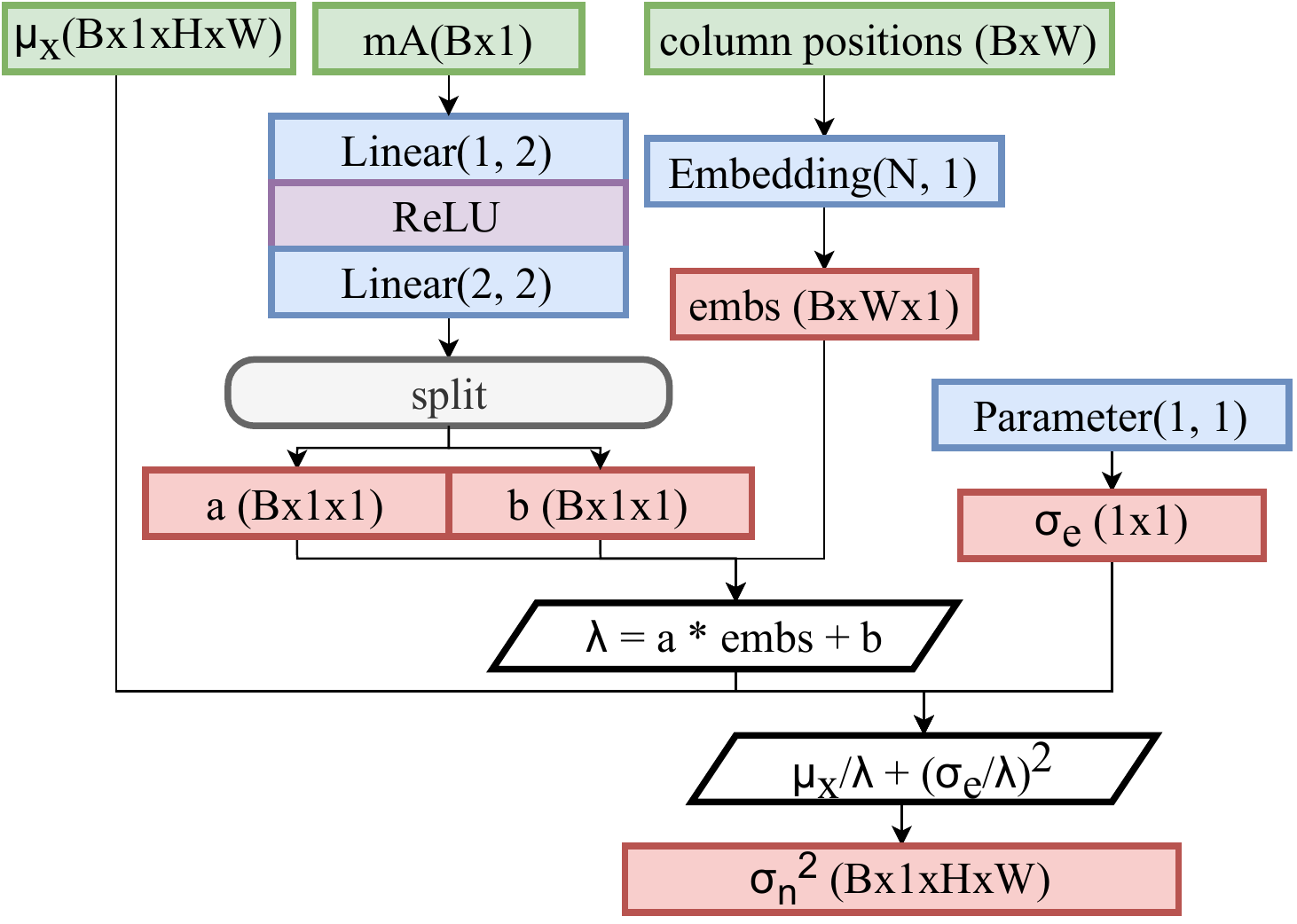}
\caption{The architecture of the noise model for self-supervised model training.}
\label{fig:arch_noise}
\end{figure}

Summarizing all the above properties, we present the noise model that allows accounting for the parameters of the acquisition process. 
Firstly, we made the model able to predict different values for different detector columns by introducing an embedding layer that maps the column number of the pixel into the value that characterizes the number of incident photons influenced by the bowtie filter. 
Then, this value is normalized by the tube current. The normalization is needed to
transform the obtained values into the distribution of incident photons for the specific current and potential.
The normalization consists of two steps. The first step is to map the tube current to the slope and bias coefficients. The mapping is made by the linear layers with $2$ output channels and ReLU activation between them. The second step is to multiply the value by the slope coefficient and add the bias coefficient. This type of normalization was chosen according to the theoretical dependency between incident flux levels of different tube currents~\cite{zeng2015simulation}. 
In this way, the proposed noise model allows predicting the approximation of the incident flux level $I_0$  depending only on the column of the pixel and acquisition parameters. There is no need to know photon distribution for the test data.
As for the electronic noise variance $\sigma_e^2$, it is represented as the parameter with the shape $1\times 1$ in the model.
The scheme of the proposed noise model is presented in Fig.~\ref{fig:arch_noise}.

If train and test projections were obtained using different CT scanner settings, for example, if the bowtie filter is removed or changed, the noise model can be tuned for the new settings and retrained on the new dataset in a self-supervised mode, and the main denoising model can be frozen during training to speed up the process. If some noise parameters ($I_0$ or $\sigma_e^2$) are known, they can be used instead of the trained ones. 

\subsection{Comparison approaches}
We decided to benchmark our approach Noise2NoisTD-ANM against the Noise2Clean, Half2Half and Noise2Void-4R approaches. The Noise2Clean model is trained using pairs of noisy and clean projections. The Half2Half approach assumes creation of pairs of noisy projections from the original projections and training the model in a supervised manner using one noisy projection as input and the other as target. The Noise2Void-4R approach is an adaptation of the approach from~\cite{laine2019hqss} for training CT projections using the proposed noise model.
All approaches used the same neural network architecture as a backbone, with some modifications caused by the requirements of the approaches.
We decided to use a relatively simple neural network that shows good results for denoising that is DnCNN (Denoising Convolutional Neural Network)~\cite{zhang2017dncnn}. In our previous work~\cite{zainulina2021n2ntd}, our model was based on the U-Net architecture~\cite{ronneberger2015u} but in this study the U-Net architecture did not show significant improvements over the DnCNN.

For the Half2Half approach, we initially created the half-dose pairs as it was described in the original paper~\cite{yuan2020half2half}. Then, we trained the model on the transmission data using MSE loss function.
Unlike the self-supervised approaches, the Noise2Clean model is trained using MSE loss function on the projection, not transmission data. Also, we used residual learning, i.e., we trained the model to predict the noise because it was shown to be effective for supervised denoising~\cite{zhang2017dncnn}. Because the noise model applied for the self-supervised approaches allows to account for the noise level depending on the used tube current automatically, we decided to include the analog of the noise level map into the input of the Noise2Clean model~\cite{zhang2018ffdnet}. 
As was shown in~\cite{zainulina2021n2ntd}, the usage of adjacent projections enhances results for the supervised training and does not give any improvements for the Noise2Void-4R approach. Therefore, we used adjacent projections as additional channels for the Noise2Clean approach and did not use them for the Noise2Void-4R approach.

\begin{figure*}[!t]
\centering
\includegraphics[width=0.8\linewidth]{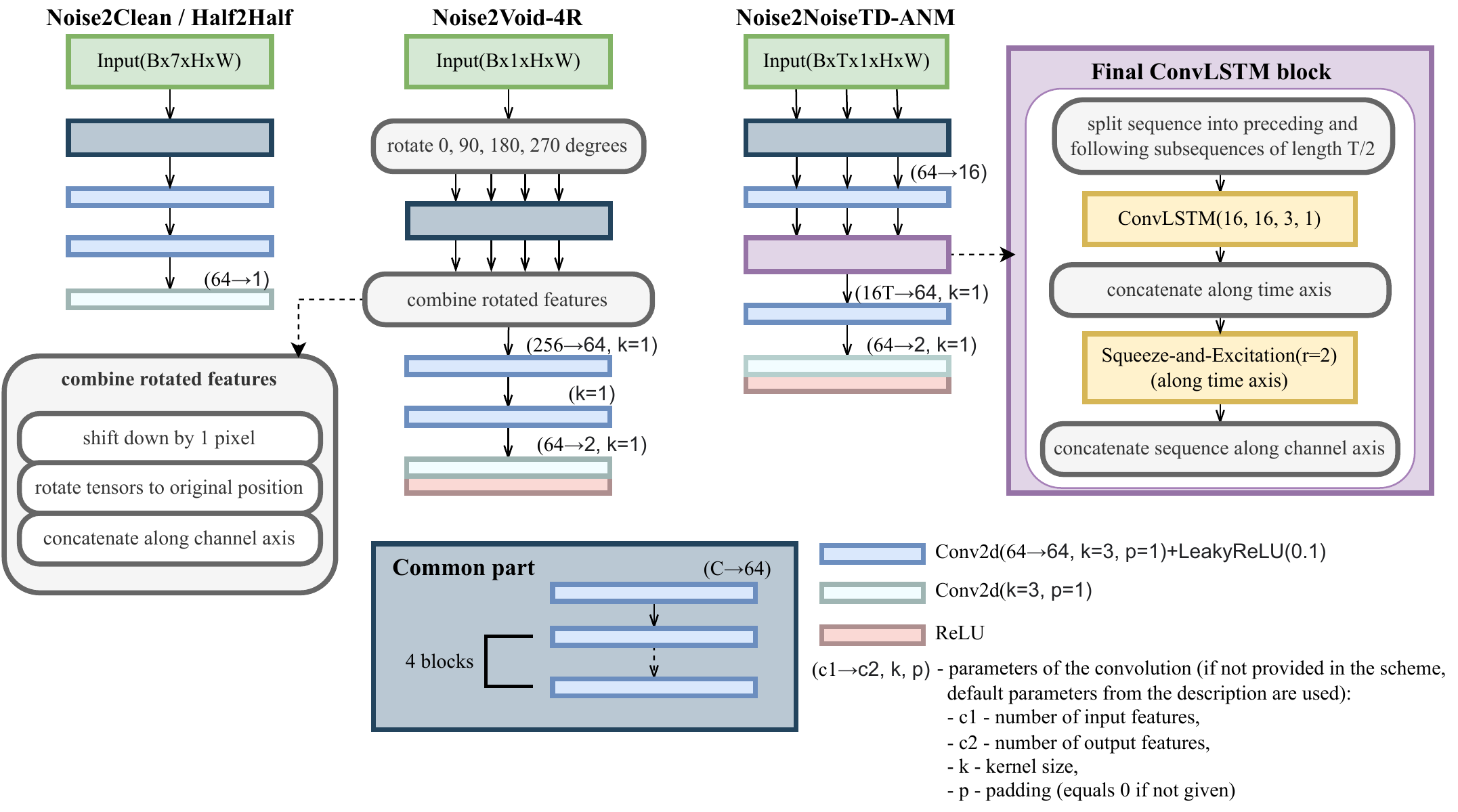}
\caption{Model architectures of the compared approaches.}
\label{fig:arch_dncnn}
\end{figure*}

The model architectures for each approach are depicted in Fig.~\ref{fig:arch_dncnn}.

\section{Experiments and results}
\subsection{Data preparation}
The study used the publicly available dataset of CT projections that is Low Dose CT Image and Projection data (LDCT-and-Projection-data)~\cite{LDCT}. The experiments were carried out on abdomen projections obtained with the fixed tube voltage $100$ kVp. During acquistion of these projections, dose modulation and a bowtie filter were used. In order to evaluate the performance of the projection-domain denoising approaches in the image domain, the reconstruction was performed by the open TIGRE toolbox~\cite{TIGRE}. Also, the dataset has information about photon distribution ($I_0$) for each CT projection that will be used for the noise model estimation.

For each patient, CT projection data are provided for both full and simulated lower dose levels. The provided low dose level is $25\%$ of the routine dose. For testing the approaches on the more severe cases of $10\%$ and $5\%$ of the routine dose, new low-dose data were simulated according to the algorithm from~\cite{zeng2015simulation}. 
Because the simulated data may differ from real data and the testing results may be different for simulated and real data, we also performed the comparison of the approaches on the data with real noise. It was possible because the dose differs for each projection due to the properties of organs and patient-specific requirements and therefore the full-dose projections having high noise levels can be found. These projections can be considered as real noisy projections. 

\subsubsection{Data selection}
For the experiments, we selected projections of several patients. The selection was based on noise levels of projections. Because noise levels of projections correlate with used tube currents (higher noise levels correspond to lower tube currents), the projections were chosen depending on the distributions of the tube currents. The noise levels of the selected projections were verified using the algorithm proposed in~\cite{liu2014noise_level} because patient anatomies can also have an impact on noise levels. 
The distribution of noise levels correlated with the distribution of the tube currents.

\begin{figure}[!t]
\centering
\includegraphics[width=2.5in]{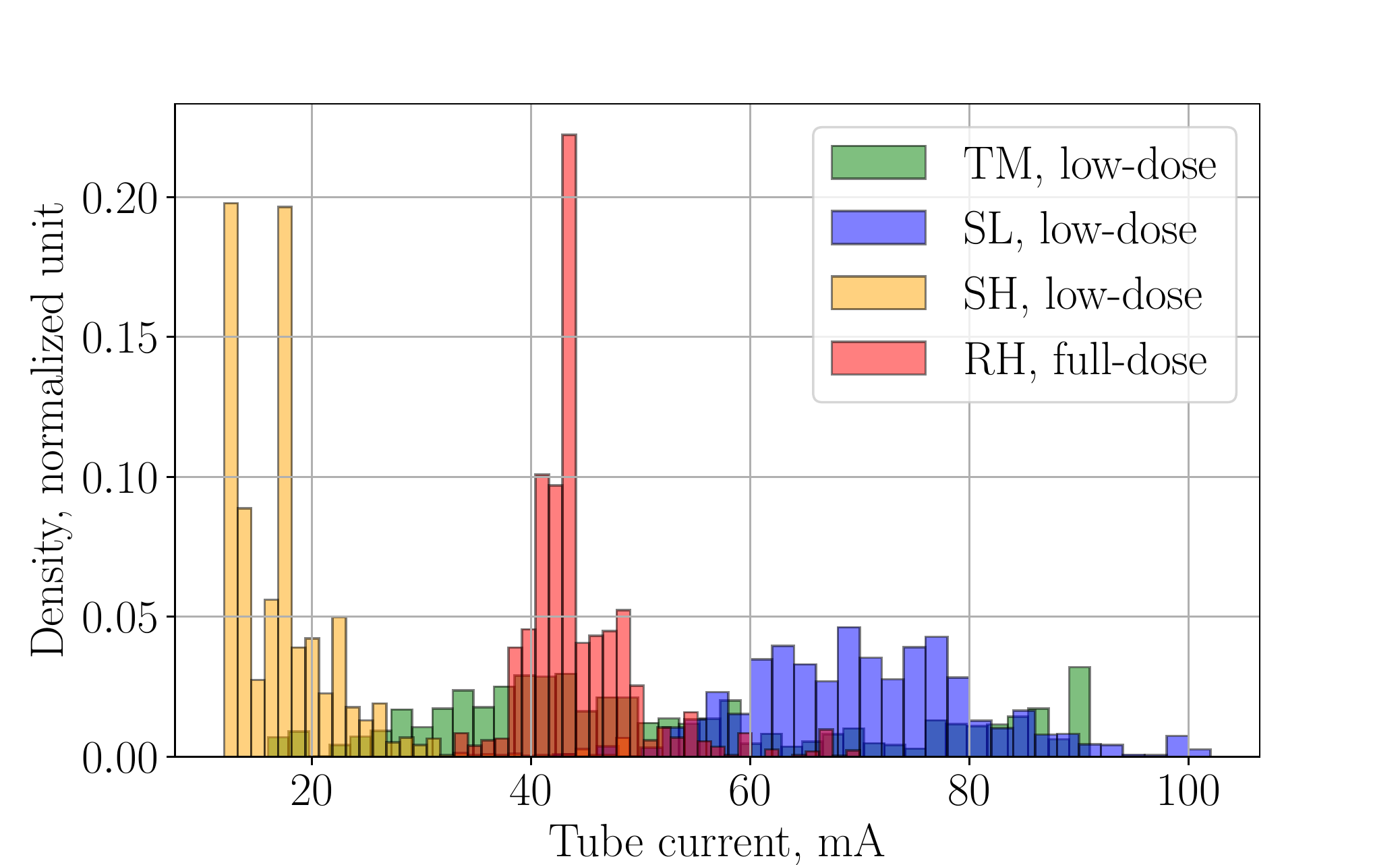}
\caption{Distribution of the tube currents (mA) used to acquire projections for different patients. The figure illustrates noise levels of the train and test sets used in the experiments because noise levels correlate with used tube currents (higher tube currents correspond to lower noise levels).}
\label{fig:hist_mAs}
\end{figure}

For the train dataset, we selected projections of one patient, whose low-dose projections have the wide range of the estimated noise levels.
We randomly picked up $21800$ projections from them so that at least $100$ projections are adjacent to be able to use the connections between neighboring frames. We denote the train dataset projections as TM (the train set, middle noise levels). Paired full-low-dose projections were used to train the Noise2Clean model, and low-dose projections were used to train the self-supervised models. For the test datasets, we chose projections of two patients with low and high noise levels of low-dose projections and picked up $6000$ consecutive projections with the highest noise levels. We denoted these sets as SL and SH, respectively, where S means that low-dose projections were simulated and the second letter shows the noise level: low and high. Their low-dose projections serve as noisy inputs to the denoising approaches, and their full-dose projections serve as reference images. Then we found full-dose projections with the highest noise level comparable to noise levels of low-dose projections of other patients. We chose $10000$ consecutive projections with the highest noise level from them to be the test dataset with real data. It was denoted as RH (real noise, high level).
Fig.~\ref{fig:hist_mAs} presents the distributions of the tube currents (mA) for the patients, whose projections were selected for the experiments. 

\subsubsection{Low-dose data simulation}
Using the algorithm from~\cite{zeng2015simulation}, we simulated low-dose projections with doses equal to $5\%$ and $10\%$ of the routine dose to test the denoising approaches on the projections with higher noise levels. We created $5\%$- and $10\%$-low-dose projections from the full-dose projections of the SL and SH sets. Also, we simulated $5\%$ TM low-dose projections to test the approaches being trained on data with the high noise level and their adaptability.

\subsection{Learning settings}
\subsubsection{Pre-estimation of the $\lambda$ parameter of the noise model}
\begin{figure}[!t]
\centering
\includegraphics[width=2.5in]{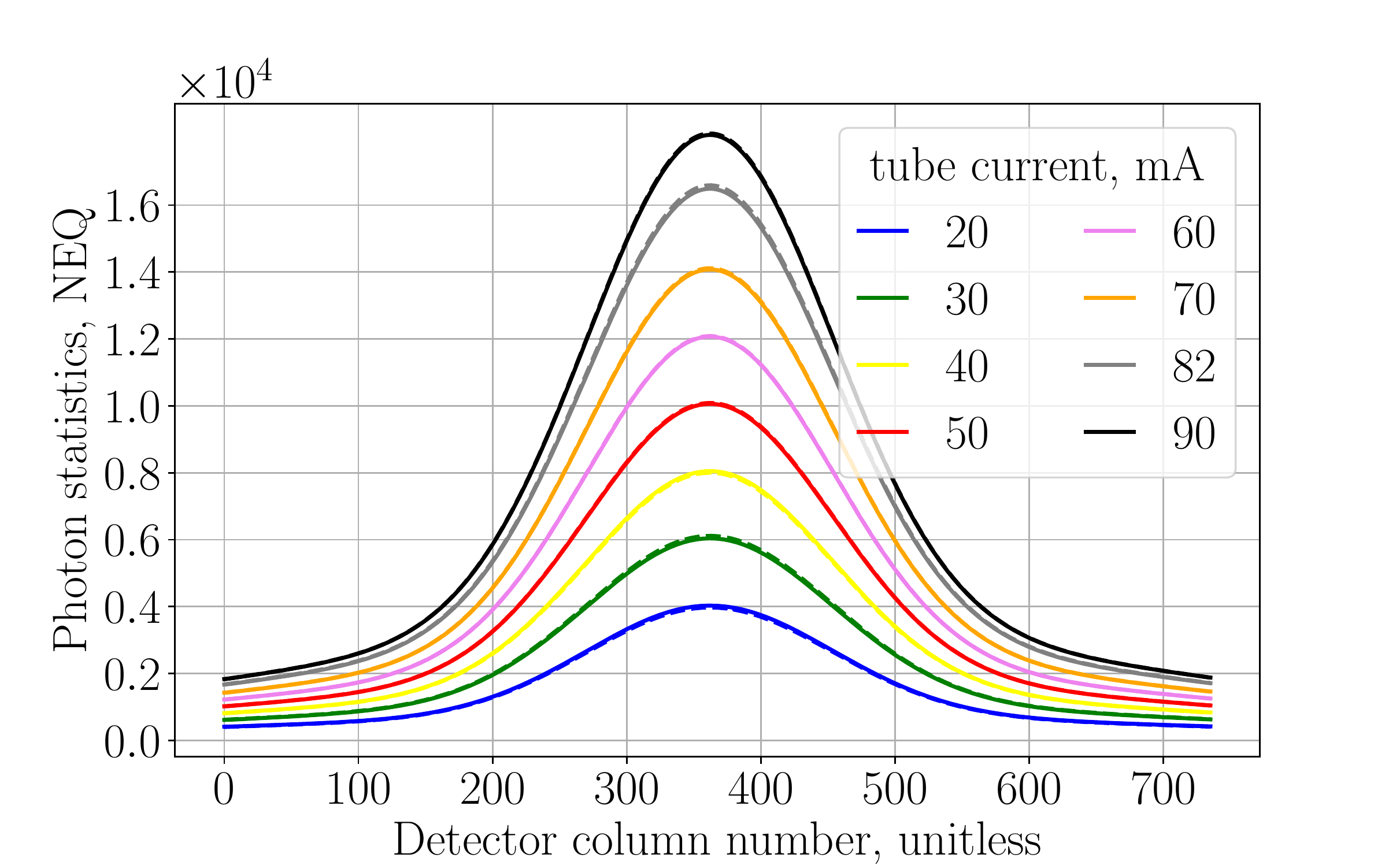}
\caption{Comparison of the predicted and actual photon distributions (Noise Equivalent Quanta) for different tube currents (mA). The solid line is predicted values, and the dashed line is true values. The figure shows that the noise estimation module successfully estimates Poisson parameter of the noise.}
\label{fig:pred_ph_dist}
\end{figure}

The noise model parameters $\lambda$ (the embedding and mapping layers) were pre-estimated using photon distributions and the information about used tube currents provided for each projection in the LDCT dataset.
We took the subset of the TM dataset and extracted information about photon distributions and corresponding values of tube currents. We trained the embedding layer and the mapping layers to predict a photon distribution from its tube current. The training was performed using the MSE loss function, Adam optimizer with learning rate $10^{-2}$ for $1000$ epochs. The optimized parameters were tested on full-dose and low-dose photon distributions and tube currents of the validation subset of the TM dataset. RMSRE (root mean square relative error)~\cite{zeng2015simulation} is equal to $0.16\pm0.11\%$ for the full-dose dataset, and it is equal to $0.53\pm0.40\%$ for the low-dose dataset. Thus, the error does not exceed $1\%$ and can be considered insignificant.
Fig.~\ref{fig:pred_ph_dist} shows predicted and original photon distributions for different tube currents.
The output values of the pre-estimated model embedding layer form a bell-shaped curve, this coincides with the theoretical shape of the photon distribution after the bowtie filter.
Then, this shape is well stretched out to fit the photon distribution corresponding to the certain tube current value.

Experiments with a joint training of the noise model and main denoising model showed that this pre-estimation makes models converge faster and give better results. Besides, freezing the pre-estimated embedding and mapping layers during joint training also leads to slightly better results. Because of that,
the pre-estimated and frozen embedding and mapping layers were used for all denoising models dependent on this noise model.

\subsubsection{Model training}
All models were trained in Pytorch using Adam optimizer with default parameters and learning rate $10^{-4}$. The minibatch used for training supervised and self-supervised models consisted of $64$ and $16$ randomly cropped $64\times 64$ patches of frames, correspondingly. The Noise2Clean model was trained using MSE loss on the projection data. The Half2Half model was trained using MSE loss on the transmission model. The Noise2Void-4R and Noise2NoiseTD-ANM models were trained using the loss function~(\ref{eq:loss}) on the transmission data. For all models, the training continued until the training curve reached a plateau. 

\subsection{Results on the simulated data}
\subsubsection{Models trained on the $25\%$-dose data}

We tested the models trained on the $25\%$-dose TM dataset on the SL and SH datasets. Low-dose projections provided by the LDCT dataset with the $25\%$ routine dose and $10\%$ and $5\%$ low-dose projections simulated by us were denoised and compared to the corresponding full-dose projections.
The quantitative comparison in the projection domain is presented in Table~\ref{tab:sim}. Additionally to the usually applied methods of evaluation of denoised images SSIM and PSNR, we used GMSD (gradient magnitude similarity deviation)~\cite{xue2013gmsd}. It considers image over-blurring unintentionally introduced by denoising, which can be overlooked by PSNR or SSIM.

The Noise2Clean approach shows the best results for the $25\%$ low-dose projections. However, the difference between the Noise2Clean and the proposed Noise2NoiseTD-ANM approach is minor, especially in the image domain, although the Noise2NoiseTD-ANM method uses less information than the supervised approach.
For the lower percents of the routine dose levels, the Noise2Clean approach gives the worse results. It shows the lowest quality for the $5\%$ low-dose projections. Although the Noise2Clean approach used the noise level map as an additional input channel, it failed to generalize to lower noise levels as the self-supervised approach did. 

As for the self-supervised approaches, the difference between them is not big. It becomes more prominent for lower doses. Although the proposed approach does not show a great advantage over other self-supervised methods in the projection domain, it gives the best results in the image domain for all doses and all image quality assessment methods.

\begin{table*}[!t]
\caption{Quantitative comparison of the denoising approaches, simulated data}
\label{tab:sim}
\centering
\begin{tabular}{c| c| c c c | c c c}\hline\hline
 &  & \multicolumn{3}{|c|}{projection domain} & \multicolumn{3}{c}{image domain}\\ \cline{3-8}
Dataset & Approach & SSIM$\uparrow$ & PSNR$\uparrow$ & GMSD$\downarrow$ & SSIM$\uparrow$ & PSNR$\uparrow$ & GMSD$\downarrow$\\
\hline\hline
\multicolumn{8}{c}{$25\%$ of the routine dose} \\\hline
& Low dose &	$0.905 \pm 0.036$ & 	$39.2 \pm 2.0$ & 	$0.013 \pm 0.006$ 
& $0.957\pm0.022$ & $43.9\pm3.0$ & $0.003\pm0.002$ \\
& Noise2Clean & $\mathbf{0.975 \pm 0.010}$ & 	$\mathbf{45.6 \pm 1.6}$ & 	$\mathbf{0.004 \pm 0.002}$ 
& $\mathbf{0.987\pm0.005}$ & $\mathbf{48.7\pm2.0}$ & $\mathbf{0.002\pm0.001}$ \\ 
SL & Half2Half & $0.973 \pm 0.010$	& $44.9 \pm 1.5$	& $0.006 \pm 0.002$ 
& $0.985\pm0.005$ & $47.3\pm1.6$ & $0.006\pm0.002$ \\
& Noise2Void-4R &	$0.973 \pm 0.010$ & 	$44.9 \pm 1.7$ & 	$0.006 \pm 0.002$	
& $0.986\pm0.005$ & $47.8\pm1.9$ & $0.004\pm0.002$ \\
& Ours &	$0.972 \pm 0.010$ & 	$44.4 \pm 1.3$ & 	$0.006 \pm 0.002$
& $\mathbf{0.987\pm0.005}$ & $48.6\pm2.1$ & $\mathbf{0.002\pm0.001}$ \\
 \hline
& Low dose &	$0.873 \pm 0.023$ & 	$37.1 \pm 1.3$ & 	$0.016 \pm 0.005$
& $0.832\pm0.030$ & $35.6\pm1.3$ & $0.015\pm0.004$ \\
& Noise2Clean &	$\mathbf{0.965 \pm 0.007}$ & 	$\mathbf{43.7 \pm 1.1}$ & 	$\mathbf{0.006 \pm 0.002}$
& $\mathbf{0.944\pm0.012}$ & $\mathbf{41.0\pm1.4}$ & $\mathbf{0.009\pm0.003}$ \\
SH & Half2Half & $0.960 \pm 0.009$	& $42.8 \pm 1.2$	& $0.010 \pm 0.003$ 
& $0.935\pm0.014$ & $39.4\pm1.8$ & $0.021\pm0.007$ \\
& Noise2Void-4R &	$0.960 \pm 0.009$ & 	$42.6 \pm 1.3$ & 	$0.010 \pm 0.004$ 
& $0.938\pm0.013$ & $39.8\pm1.7$ & $0.016\pm0.005$ \\
& Ours &	$0.961 \pm 0.008$ & 	$42.3 \pm 1.0$ & 	$0.009 \pm 0.004$ 
& $0.941\pm0.011$ & $40.5\pm1.4$ & $\mathbf{0.009\pm0.003}$ \\
 \hline\hline
\multicolumn{8}{c}{$10\%$ of the routine dose} \\\hline
& Low dose &	$0.787 \pm 0.075$ & 	$33.9 \pm 2.8$ & 	$0.044 \pm 0.025$ 
& $0.845\pm0.103$ & $37.3\pm5.0$ & $0.024\pm0.024$ \\
& Noise2Clean &	$0.958 \pm 0.029$ & 	$41.6 \pm 4.3$ & 	$0.023 \pm 0.023$ 
& $0.939\pm0.057$ & $42.5\pm5.7$ & $0.016\pm0.017$ \\
SL & Half2Half & $0.957 \pm 0.018$	& $42.7 \pm 1.9$	& $0.011 \pm 0.004$ 
& $0.979\pm0.008$ & $45.8\pm1.8$ & $0.008\pm0.002$ \\
& Noise2Void-4R &	$0.961 \pm 0.015$ & 	$\mathbf{43.1 \pm 1.6}$ & 	$0.010 \pm 0.003$ 
& $0.980\pm0.007$ & $46.1\pm1.9$ & $0.007\pm0.002$ \\
& Ours &	$\mathbf{0.963 \pm 0.010}$ & 	$42.8 \pm 1.0$ & 	$\mathbf{0.007 \pm 0.002}$ 
& $\mathbf{0.981\pm0.006}$ & $\mathbf{46.7\pm1.8}$ & $\mathbf{0.004\pm0.002}$ \\
\hline
& Low dose &	$0.860 \pm 0.026$ & 	$36.0 \pm 1.9$ & 	$0.023 \pm 0.012$
& $0.813\pm0.037$ & $34.6\pm1.9$ & $0.021\pm0.013$ \\
& Noise2Clean &	$\mathbf{0.963 \pm 0.012}$ & 	$42.5 \pm 3.2$ & 	$0.012 \pm 0.015$ 
& $0.938\pm0.020$ & $40.2\pm2.3$ & $0.013\pm0.009$ \\
SH & Half2Half & $0.962 \pm 0.008$	& $43.0\pm 1.2$	& $0.010 \pm 0.003$ 
& $0.938\pm0.013$ & $39.5\pm1.8$ & $0.022\pm0.007$ \\
& Noise2Void-4R &	$0.962 \pm 0.008$ & 	$\mathbf{43.0 \pm 1.1}$ & 	$0.011 \pm 0.003$ 
& $0.935\pm0.015$ & $39.2\pm1.9$ & $0.024\pm0.008$ \\
& Ours &	$0.961 \pm 0.009$ & 	$41.4 \pm 0.8$ & 	$\mathbf{0.009 \pm 0.003}$ 
& $\mathbf{0.942\pm0.013}$ & $\mathbf{40.4\pm1.6}$ & $\mathbf{0.012\pm0.004}$ \\
 \hline\hline
\multicolumn{8}{c}{$5\%$ of the routine dose} \\\hline
& Low dose &	$0.632 \pm 0.106$ & 	$28.2 \pm 3.3$ & 	$0.108 \pm 0.050$
& $0.565\pm0.262$ & $28.1\pm7.2$ & $0.113\pm0.081$ \\
& Noise2Clean &	$0.863 \pm 0.103$ & 	$31.6 \pm 5.9$ & 	$0.099 \pm 0.059$ 
& $0.707\pm0.228$ & $32.5\pm7.9$ & $0.086\pm0.062$ \\
SL & Half2Half & $0.918 \pm 0.039$	& $39.2 \pm 2.4$	& $0.025 \pm 0.012$
& $0.961\pm0.019$ & $43.5\pm2.4$ & $0.010\pm0.003$ \\
& Noise2Void-4R &	$0.934 \pm 0.027$ & 	$40.5 \pm 1.9$ & 	$0.017 \pm 0.004$ 
& $0.968\pm0.014$ & $44.0\pm2.2$ & $0.011\pm0.004$ \\
& Ours &	$\mathbf{0.963 \pm 0.012}$ & 	$\mathbf{41.9 \pm 1.1}$ & 	$\mathbf{0.011 \pm 0.003}$ 
& $\mathbf{0.980\pm0.007}$ & $\mathbf{45.9\pm1.8}$ & $\mathbf{0.007\pm0.003}$ \\
 \hline
& Low dose &	$0.742 \pm 0.055$ & 	$31.1 \pm 2.9$ & 	$0.058 \pm 0.028$ 
& $0.569\pm0.118$ & $27.5\pm3.5$ & $0.087\pm0.046$ \\
& Noise2Clean &	$0.935 \pm 0.034$ & 	$36.3 \pm 5.9$ & 	$0.048 \pm 0.038$
& $0.779\pm0.123$ & $32.3\pm4.5$ & $0.060\pm0.037$ \\
SH & Half2Half & $0.944 \pm 0.016$	& $40.8 \pm 1.7$	& $0.017 \pm 0.006$
& $0.916\pm0.020$ & $38.1\pm1.8$ & $0.027\pm0.008$ \\
& Noise2Void-4R &	$\mathbf{0.956 \pm 0.009}$ & 	$\mathbf{41.9 \pm 1.4}$ & 	$0.016 \pm 0.006$ 
& $0.924\pm0.017$ & $38.1\pm2.1$ & $0.033\pm0.011$ \\
& Ours &	$\mathbf{0.956 \pm 0.009}$ & 	$40.7 \pm 0.8$ & 	$\mathbf{0.012 \pm 0.004}$ 
& $\mathbf{0.930\pm0.015}$ & $\mathbf{39.2\pm1.7}$ & $\mathbf{0.019\pm0.006}$ \\
\hline\hline
\end{tabular}
\end{table*}

\begin{figure*}[!t]
\centering
\subfloat[from the SL dataset]{\includegraphics[width=0.5\linewidth]{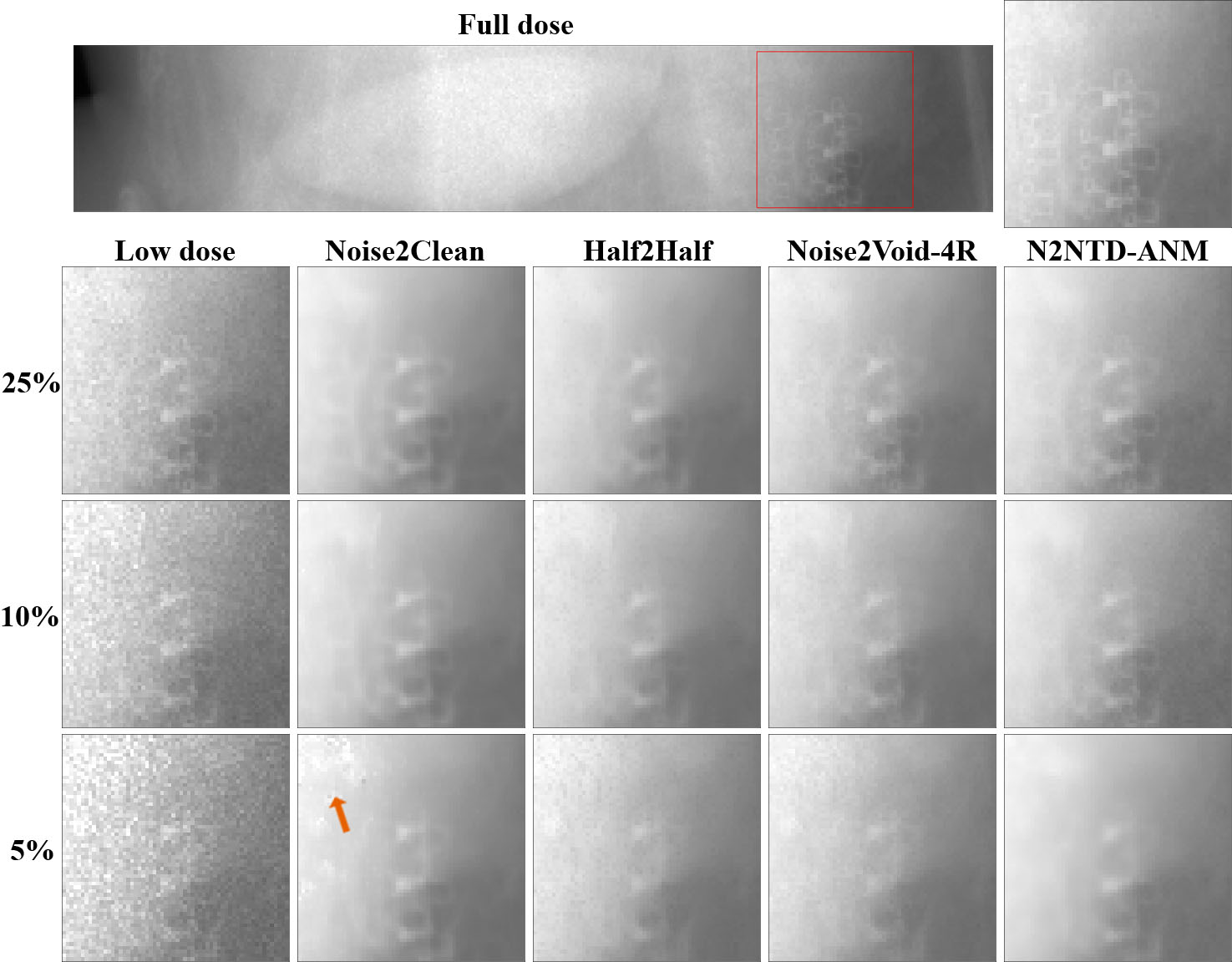}%
}
\hfil
\subfloat[from the SH dataset]{\includegraphics[width=0.5\linewidth]{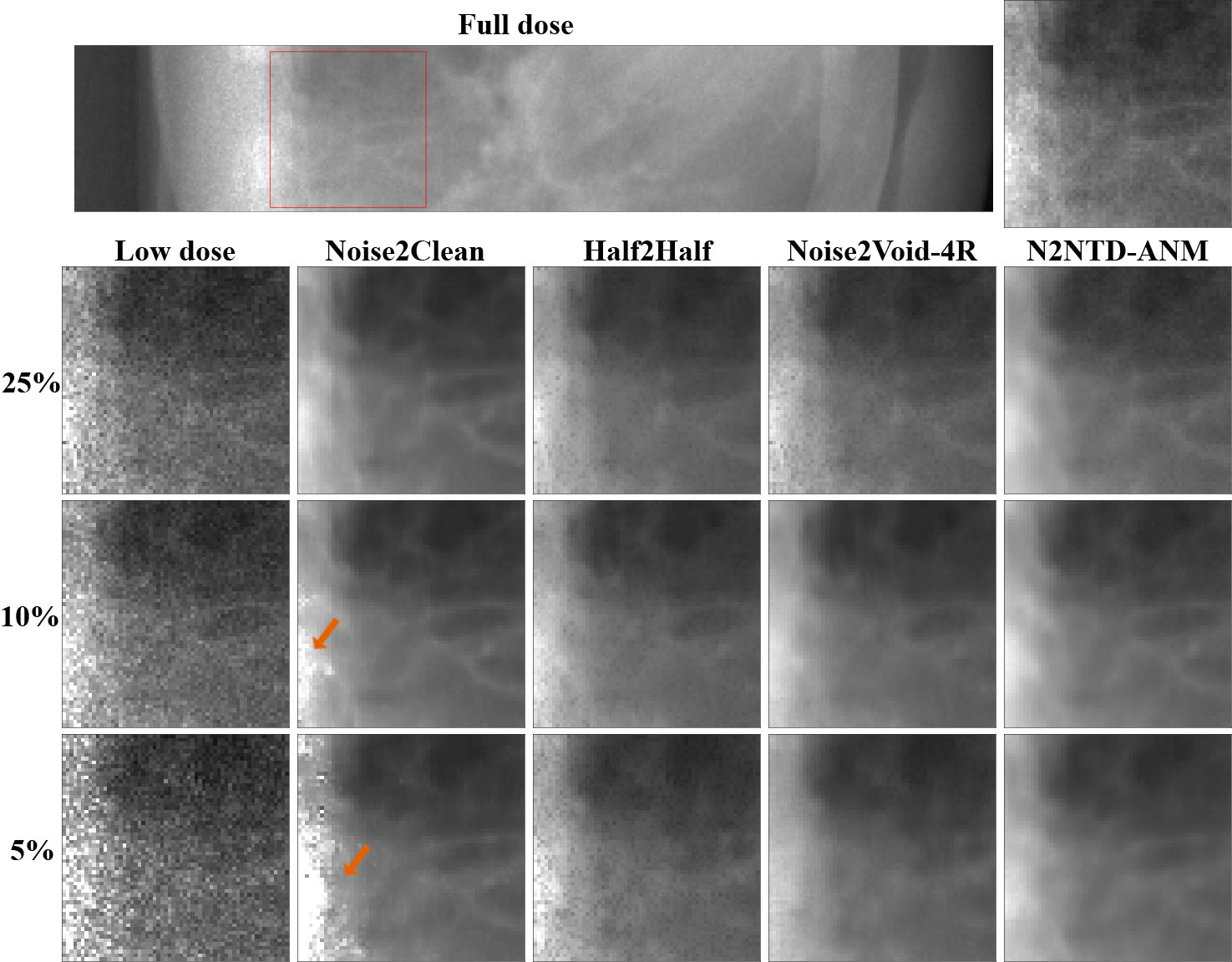}%
}
\caption{The example parts of CT projections of different doses ($25\%$, $10\%$, $5\%$) denoised by the approaches. N2NTD-ANM stands for Noise2NoiseTD-ANM. The arrow points to the parts that the Noise2Clean model failed to restore due to high noise levels.}
\label{fig:sim_prj}
\end{figure*}

\begin{figure*}[!t]
\centering
\includegraphics[width=\linewidth]{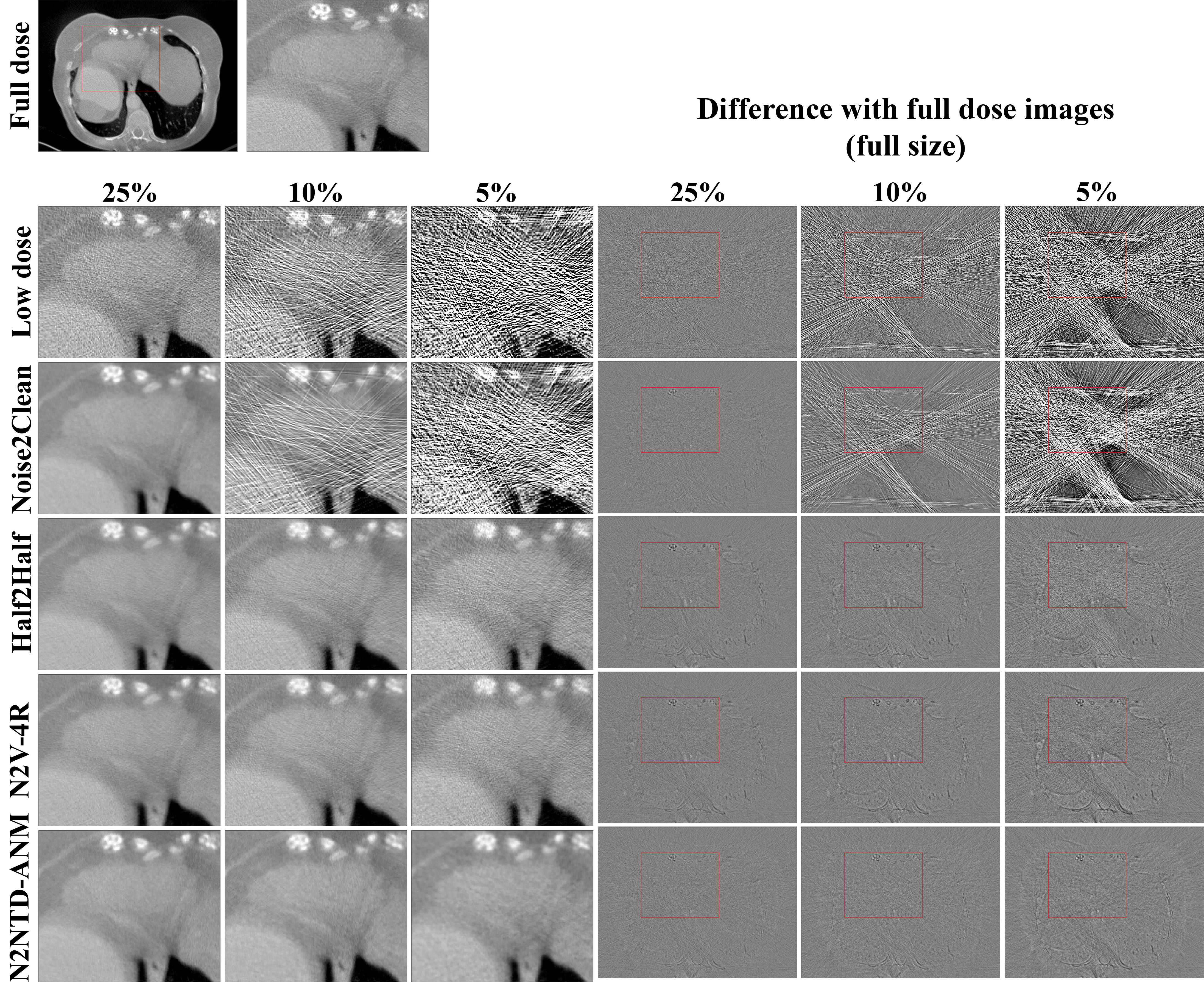}
\caption{The comparison of CT slices reconstructed from the denoised $25\%$, $10\%$, $5\%$ low-dose projections (from the SL dataset). N2V-4R and N2NTD-ANM stand for Noise2Void-4R and Nois2NoiseTD-ANM, respectively.}
\label{fig:sim_img}
\end{figure*}

The approaches were also evaluated qualitatively. Fig.~\ref{fig:sim_prj} shows the example parts of the denoised low-dose CT projections. The CT slices reconstructed from these projections are presented in Fig.~\ref{fig:sim_img}. It can be found from the figures that although the Noise2Clean approach preserves structural details rather good, being trained on the data with lower noise levels, it fails to restore pixels corrupted by more severe noise. This led to streaks and noise on the reconstructed CT slices.
At the same time, the $25\%$ low-dose projections denoised by the Noise2NoiseTD-ANM and the reconstructed from them CT slices have the quality comparable to the CT projections and the reconstructed slices of the Noise2Clean approach.
The figures also confirm that the Noise2NoiseTD-ANM approach preserves edges and small details better than the Noise2Void-4R and Half2Half models. 

\subsubsection{Cross-test}
\begin{figure*}[!t]
\centering
\subfloat{\includegraphics[width=0.45\linewidth]{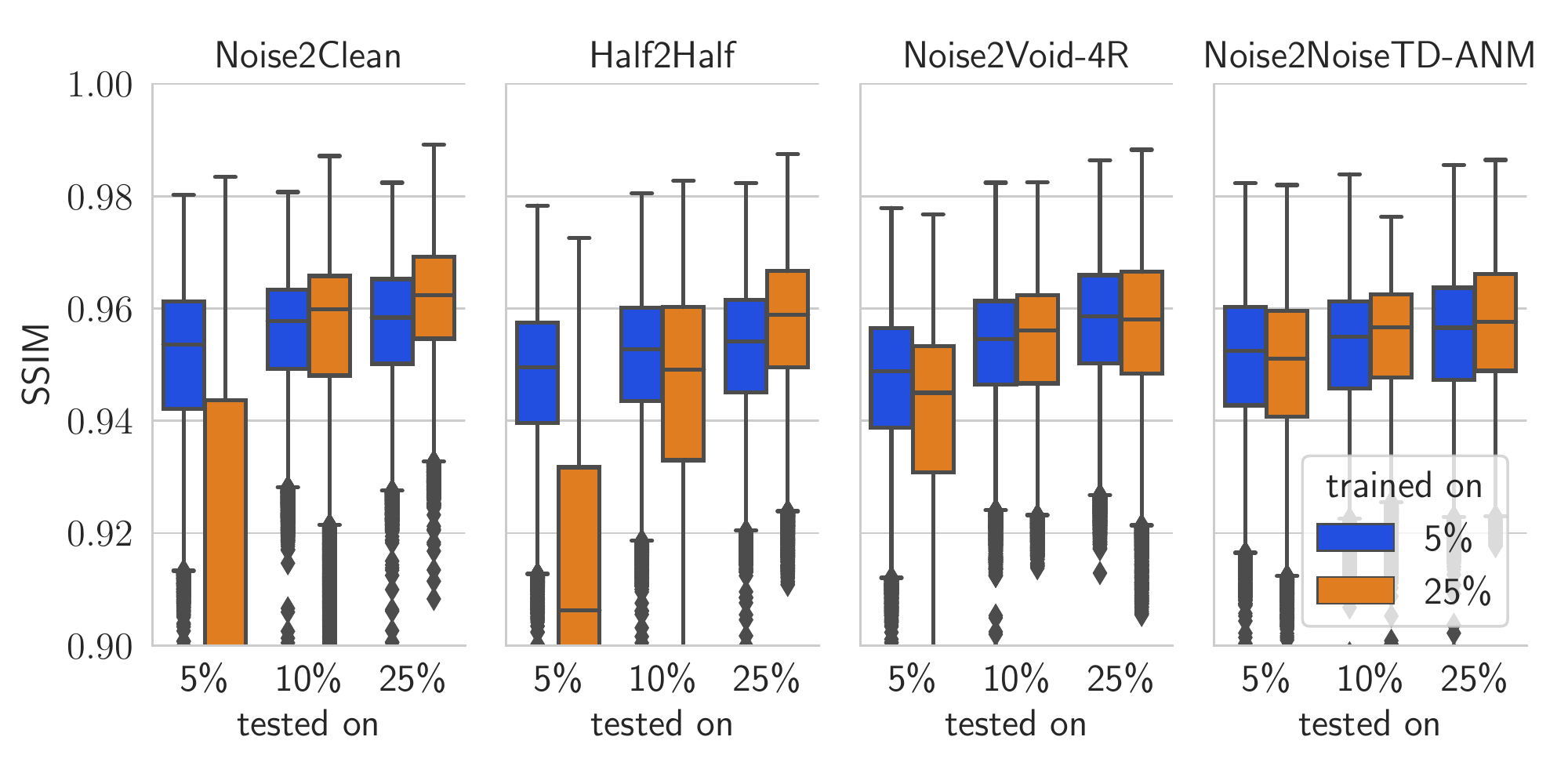}%
}
\hfil
\subfloat{\includegraphics[width=0.45\linewidth]{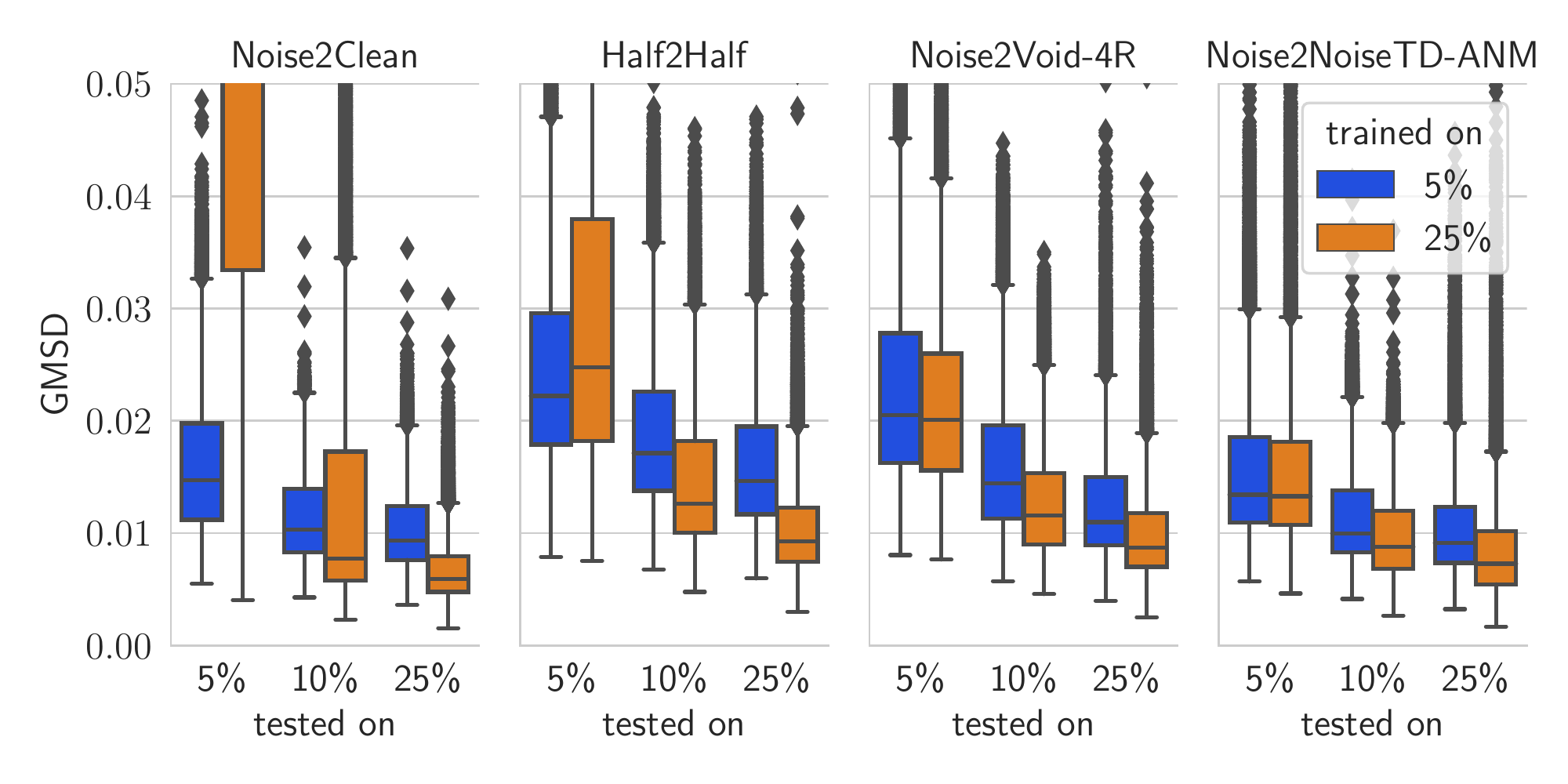}%
}
\caption{The quantitative comparison of the denoising models trained on $25\%$ and $5\%$ low-dose projections using box plots in the projection domain.}
\label{fig:cross_test}
\end{figure*}

For testing the capability of the approaches to make a neural network able to generalize to different noise levels, we performed cross-testing. Additionally to the models trained on the $25\%$-dose TM dataset, we trained the models of each approach on the $5\%$-dose TM datasets using the same training setting. The trained models were tested on the five sets of projections ($6000$ projections in each set) including the SL and SH datasets. Three other datasets were selected such as their noise levels were between the noise levels of the SL and SH datasets. Each testing dataset was made at $25\%$, $10\%$, and $5\%$ of the routine dose. The results of the cross-testing are presented in Fig.~\ref{fig:cross_test}.

The box plots show that the difference between the Noise2Clean models trained on the datasets made at $25\%$ and $5\%$ doses is significant. The difference is less between the Half2Half models but it is more pronounced than for the Noise2Void-4R and Noise2NoiseTD-ANM models. Thus, the Noise2NoiseTD-ANM and Noise2Void-4R models generalize better to various noise levels. They show approximately the same tendency in the differences between models trained at dose levels of $25\%$ and $5\%$. The both methods used the proposed noise model that improved their generalization capability.
The advantage of the noise model in making the denoising adaptable to various noise levels will also be further proven by comparing between using the noise model and the simple MSE-loss function on the projection data for training.

In the comparison of the self-supervised models, the Noise2NoiseTD-ANM model shows slightly better results, this is more pronounced on the GMSD box plot and may become more visible in the image domain. Although the difference between the approaches is not prominent for $25\%$ of the routine dose, it increases with the decrease of the dose and becomes clear for $5\%$ of the routine dose.
The advantage of the Noise2NoiseTD-ANM model over the Noise2Void-4R model is that it leverages information from adjacent projections that helps to preserves edges and some small details. The Noise2Void-4R restores a pixel depending on the neighborhood of this pixel from one image to be denoised. If this neighborhood contains a big fraction of the corrupted pixels, it is less likely to restore the pixel correctly. At the same time, the usage of the adjacent projections by the Noise2NoiseTD-ANM approach decreases the error.

\subsection{Results on the real data}
We tested the Noise2Clean and Noise2NoiseTD-ANM approaches on the RH dataset. 
Because the proposed method assumes training without high-quality reference images, we also trained the Noise2NoiseTD-ANM model directly on the RH dataset with the same learning settings. We denoted this model as Noise2NoiseTD-ANM$^*$.
The comparison was performed only qualitatively because there are no high-quality images for assessment of the methods using full-reference quality measures and there are no reference-free measures proven to be representative. 

\begin{figure}[!t]
\centering
\includegraphics[width=\linewidth]{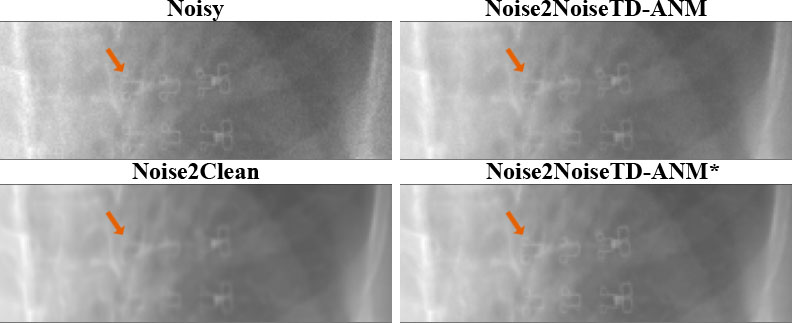}
\caption{The comparison of the parts of the denoised CT projections with real noise. The arrow shows that the Noise2NoiseTD-ANM models better preserve structures than the Noise2Clean model.}
\label{fig:real_prj}
\end{figure}

\begin{figure}[!t]
\centering
\includegraphics[width=\linewidth]{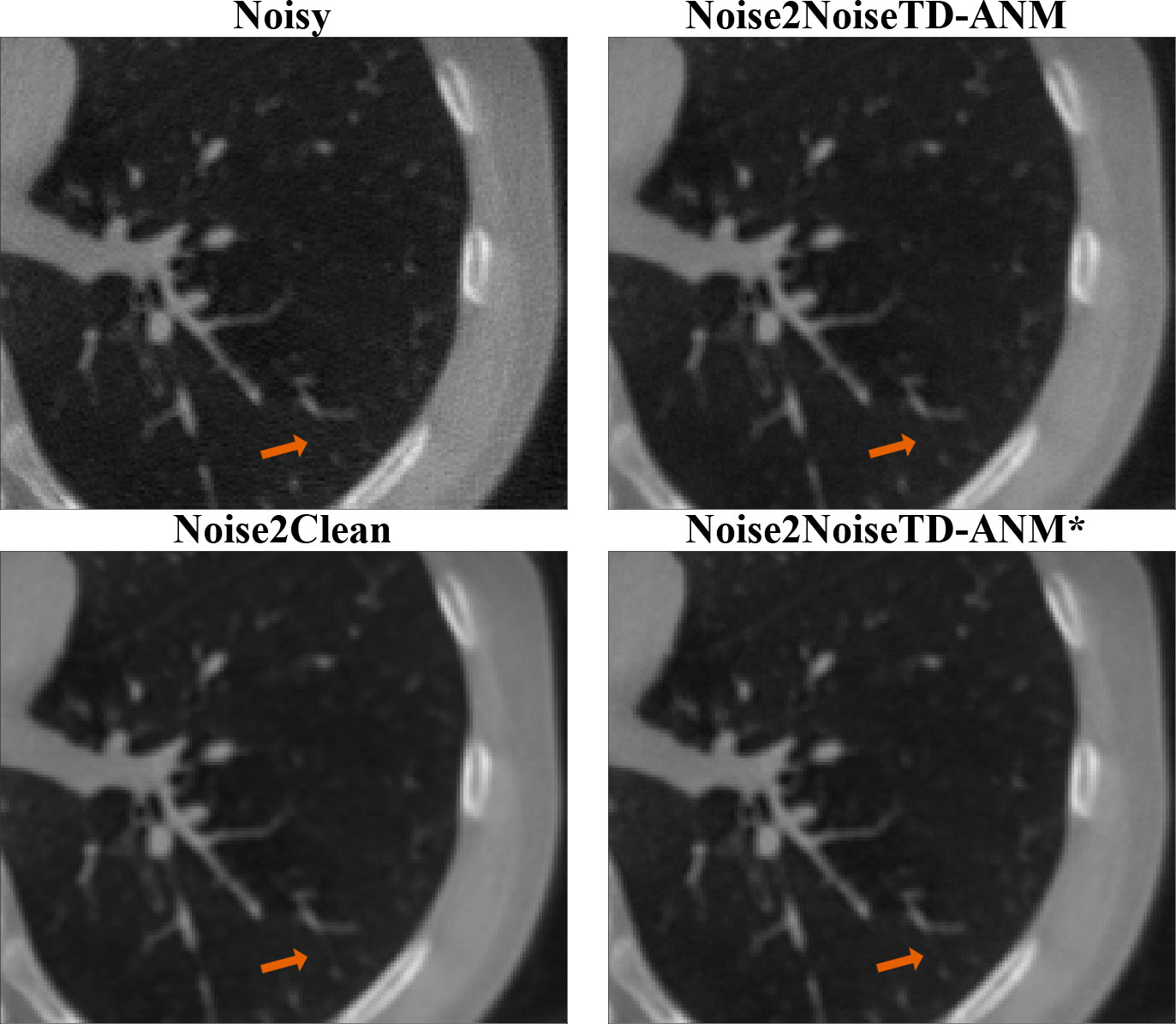}
\caption{The comparison of the example parts of CT slices reconstructed from the denoised projections that had real noise. The arrow shows that the Noise2Clean model almost connected structures that were originally disconnected, unlike the Noise2NoiseTD-ANM models, due to stronger smoothing.}
\label{fig:real_img}
\end{figure}

The comparison is presented in Fig.~\ref{fig:real_prj}. The figure demonstrates that the Noise2NoiseTD-ANM model removed the noise slightly worse than the Noise2Clean, but it kept some structural details better. The Noise2NoiseTD-ANM$^*$ model trained on the data with real noise suppressed the noise better than the Noise2NoiseTD-AMNM model while maintaining the same level of detail preservation.
The parts of the CT slices reconstructed from the denoised projections are shown in Fig.~\ref{fig:real_img}. The orange row indicates the small area in the lung where for the Noise2Clean model the light band occurred as a result of over-smoothing while it does not exist in the original noisy image and did not appear for the Noise2NoiseTD-ANM models.

The Noise2NoiseTD-ANM$^*$ model demonstrated higher visual quality. Thus, the model training directly on the data with real noise can be advantageous and the application of the Noise2NoiseTD-ANM approach on the real data can result in more accurate denoising than the application of the Noise2Clean model trained on the simulated data.

\subsection{Experiments with the noise model}
\begin{figure}[!t]
\centering
\includegraphics[width=\linewidth]{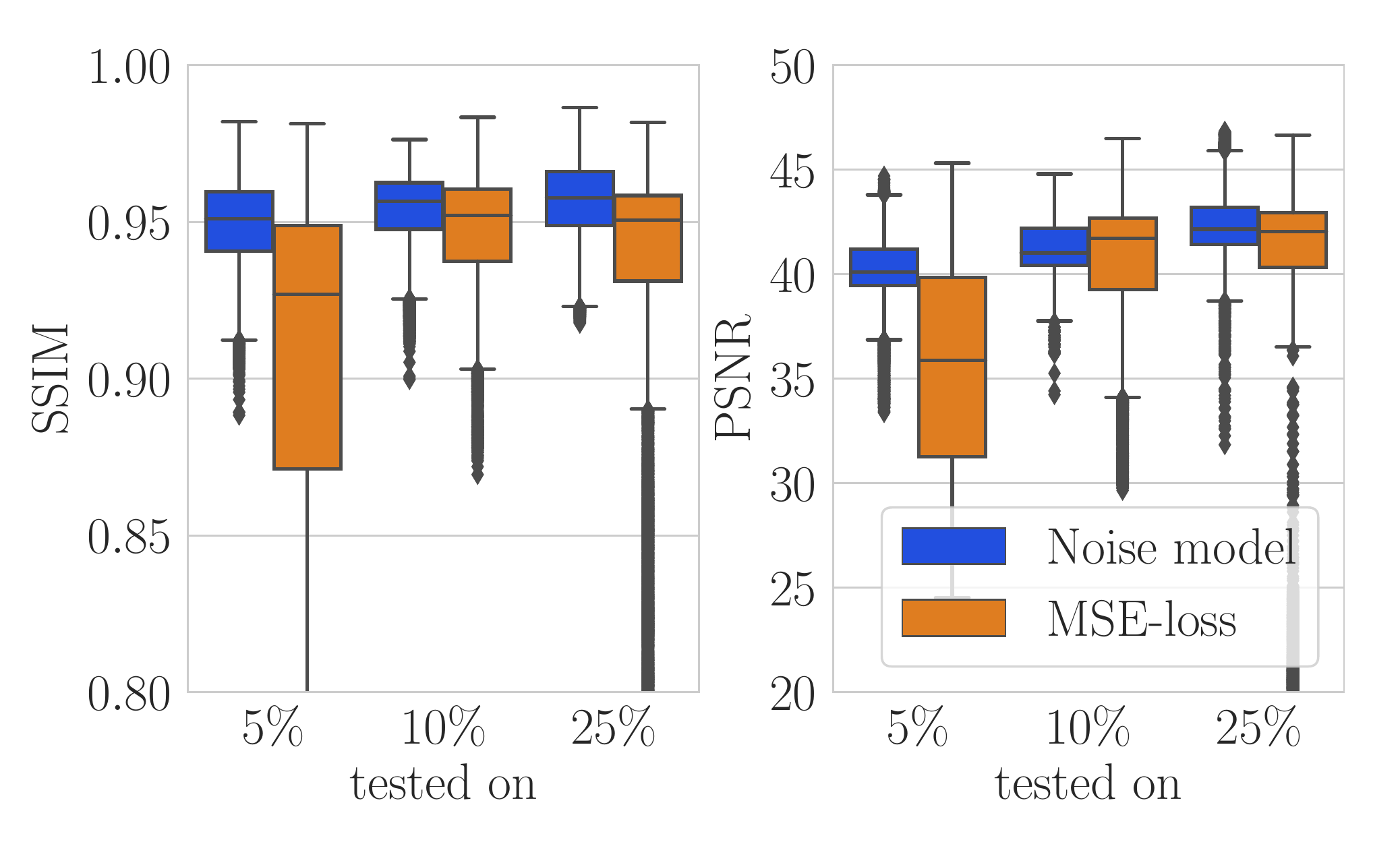}
\caption{The quantitative comparison of the noise model and the MSE-loss function in the projection domain.}
\label{fig:mse_vs_noise}
\end{figure}

The experiment with cross-testing showed that the noise model could help the self-supervised Noise2Void-4R and Noise2NoiseTD-ANM models better perform on the data with unseen noise levels.
To ensure that the better adaptation of the self-supervised models to different noise levels is the advantage of the used noise model we tested the used noise model against the simple MSE loss function. We trained the Noise2NoiseTD-ANM model (without ReLU activation at the end) using the MSE loss function on the projection data using the earlier defined learning settings. The model trained using the MSE-loss function was compared to the Noise2NoiseTD-ANM model trained using the noise model on the five test datasets used for the cross-testing in the projection domain.
The results (Fig.~\ref{fig:mse_vs_noise}) prove that the proposed noise model enhances generalization capabilities of the denoising model. 

\begin{figure}[!t]
\centering
\subfloat[parts of CT projections]{\includegraphics[width=\linewidth]{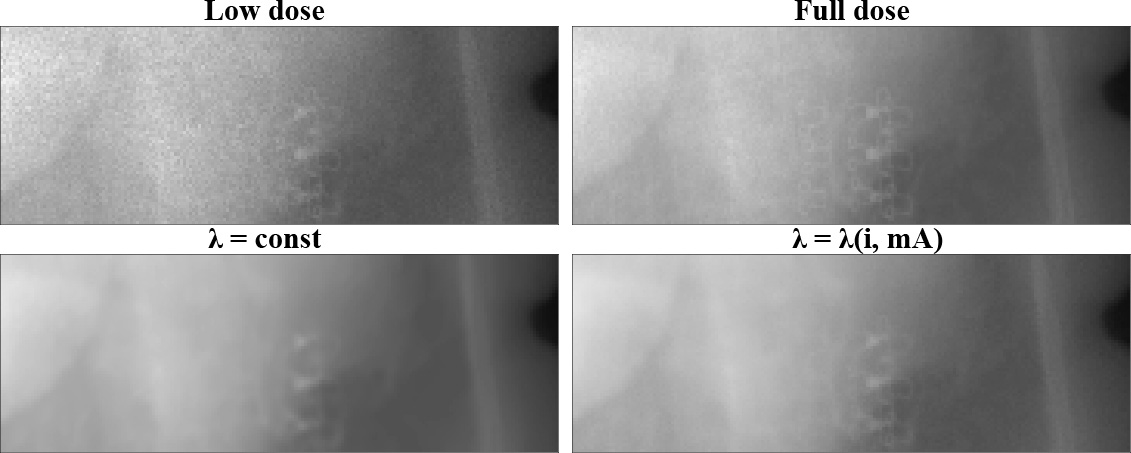}}
\vfil
\subfloat[parts of CT slices]{\includegraphics[width=0.95\linewidth]{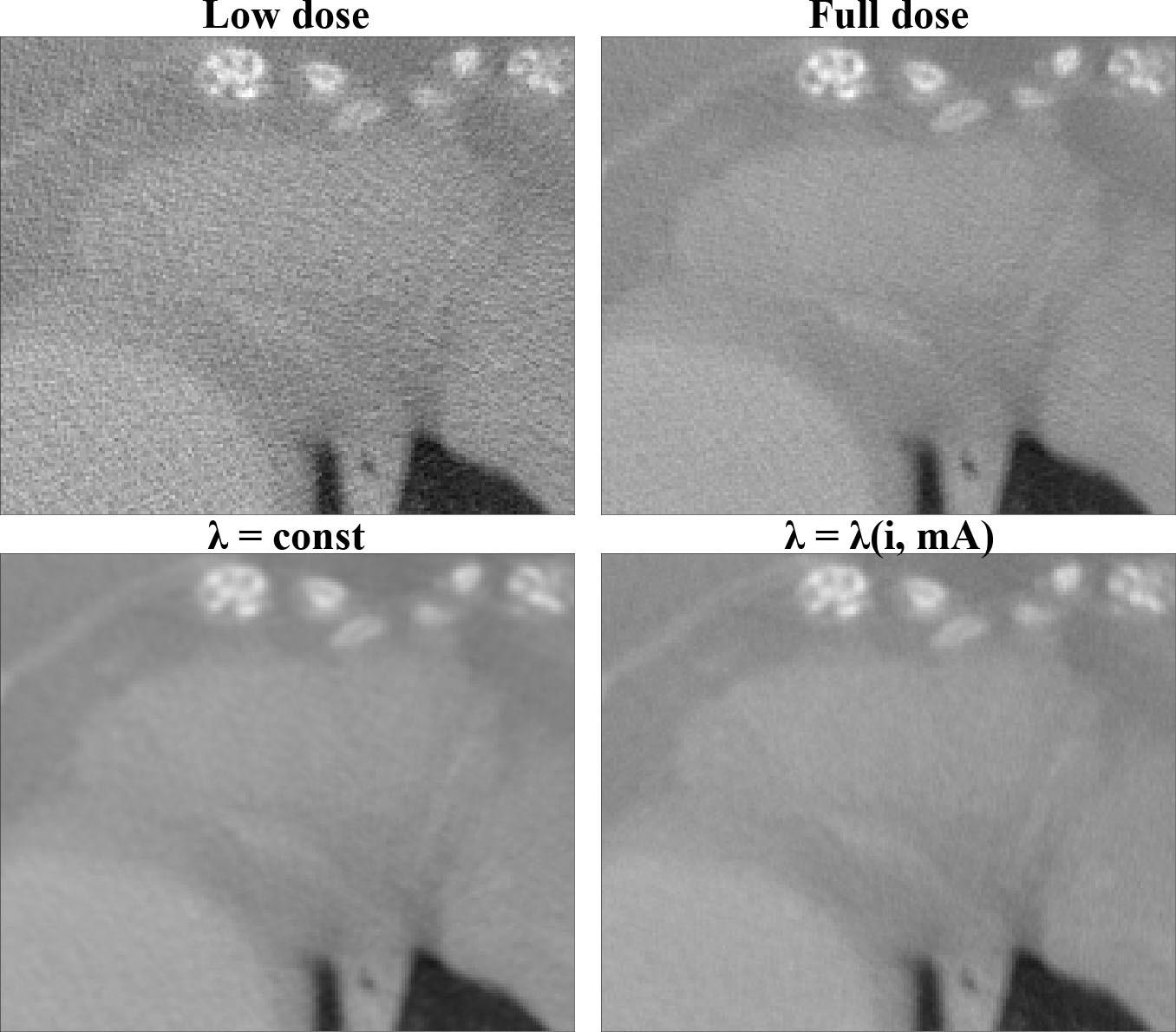}
}
\caption{The example parts of CT projections from the SL dataset denoised by the Noise2NoiseTD-ANM with different noise models and the CT slices reconstructed from them.}
\label{fig:noise_prj}
\end{figure}

Additionally, we tested whether considering CT scanner properties gives improvements. We compared the proposed noise model ($\lambda=\lambda(i, mA)$) with the model that assumes Gaussian approximation of the mixed Poisson-Gaussian distribution of the noise with the constant parameters $\lambda$ ($\lambda=const$) and $\sigma_e^2$ of the noise variance. These parameters can hardly be pre-estimated in this case. 
For comparison, we trained the Noise2NoiseTD-ANM model in the transmission domain with the noise variance's $\lambda$ parameter depending on neither a pixel position nor the tube current ($\lambda=const$). 
The results of the comparison in the projection domain are presented in Table~\ref{tab:noise}. Fig.~\ref{fig:noise_prj} gives the visual comparison of the example denoised projection parts and the obtained CT slices.
The results show that varying parameter $\lambda$ of the noise variance allows models to remove noise better while preserving edges and fine details. Besides, the independence of the parameter $\lambda$ from the tube current would not allow the models easily adapt to different noise levels, the parameter would need to be retrained or it would worsen the results.

\begin{table}[!t]
\caption{The quantitative evaluation of the noise model with the Poisson parameter $\lambda$ dependent on the detector column position and tube current ($\lambda=\lambda(i, mA)$) and the constant Poisson parameter $\lambda$ ($\lambda=const$) in the projection domain.}
\label{tab:noise}
\centering
\begin{tabular}{c| c c c}\hline\hline
Noise model & SSIM$\uparrow$ & PSNR$\uparrow$ & GMSD$\downarrow$ \\ 
\hline\hline
\multicolumn{4}{c}{SL dataset}\\\hline
$\lambda=const$ &	$0.967 \pm 0.011$ & 	$44.0 \pm 1.3$ & 	$0.008 \pm 0.002$ \\ 
$\lambda=\lambda(i, mA)$ &	$0.972 \pm 0.010$ & 	$44.4 \pm 1.3$ & 	$0.006 \pm 0.002$ \\
\hline\hline 
\multicolumn{4}{c}{SH dataset}\\\hline
$\lambda=const$ &	$0.956 \pm 0.009$ & 	$42.5 \pm 1.2$ & 	$0.010 \pm 0.004$  \\ 
$\lambda=\lambda(i, mA)$ &	$0.961 \pm 0.008$ & 	$42.3 \pm 1.0$ & 	$0.009 \pm 0.004$  \\ \hline

\hline\hline
\end{tabular}
\end{table}

\section{Discussion and conclusions}
In the study, we considered self-supervised deep-learning methods for denoising low-dose CT projections. Although supervised methods were shown to perform better than the self-supervised approaches for the denoising task, they require the representative dataset of paired clean-noisy images, which is difficult to obtain in the clinical setting. 
As for synthetic approaches (e.g., using GANs \cite{prokopenko2019unpaired}), it is not always possible to obtain the complete information about the CT machine necessary for the accurate simulation. Besides, a closer proximity of the distributions of training and testing data is known to lead to better results, making the training on real data preferable.
Therefore, self-supervised approaches for denoising low-dose CT images are of particular value. Early self-supervised approaches mostly built upon the Noise2Noise model, with the generation of the dataset complicating the denoising challenge and making the solution less adaptable.

Herein, we proposed a new self-supervised approach Noise2NoiseTD-ANM that uses only the original noisy projections. This method is based on restoring a certain CT projection from its adjacent projections, as in the problem of frame prediction from the frame sequence, and modeling data distributions for model training and inference. 
For this, we included ConvLSTM units in the network and developed the noise model that takes into account the theoretical physics-based noise distribution of the CT projections. This noise model incorporates the effect of bowtie filtering, adapting to the dose modulation.
It should be noted that the train-inference scheme was taken from~\cite{laine2019hqss} and adopted for denoising CT projections using the developed noise model.  

Using the same backbone neural network architecture, we compared our approach with the Noise2Clean, the Half2Half, and the state-of-the-art blind-spot network-based approach~\cite{laine2019hqss}, using the adapted train-inference scheme.  We tested the models on the simulated test data with both high and low noise levels. The results showed that in the case of training and testing on the data with approximately the same noise levels, the Noise2Clean approach gives slightly better results. However, the Noise2Clean model, even with given noise level maps, fails to properly denoise projections of different noise levels, whereas, our physics-based self-supervised approach with realistic noise model is more successful in generalizing to any noise levels. These experiments also emphasize the advantage of the self-supervised technique over the supervised training regime. 
The noise model, capable of estimating the noise properties separately, also allows to easily adapt to various denoising scenarios. For example, if the test noise distribution model differs from the noise model of the trained model, it can be easily tuned, and then the main denoising model and the noise model can be retrained together on the new data, or only the parameters of the noise model can be optimized.

In the comparison of the self-supervised approaches, our approach marginally outperformed the others. While the approach from~\cite{laine2019hqss} relies only on the neighborhood of a pixel to be denoised, our approach looks at the adjacent projections that contain almost the same content but different noise realization. Therefore, our approach is less prone to losing some object details, even in the cases of severe noise. 

Besides, we compared the approaches on test projections with real noise. Even trained on the simulated data, the proposed approach turned out to be better than the supervised approach. Nevertheless, our approach can be directly trained on the data to be denoised. As was expected, training of the Noise2NoiseTD-ANM model on the test data with real noise gave improvements. However, the comparison was performed only qualitatively. Similarly to other medical imaging modalities, the quantitative comparison could not be performed because of the lack of representative no-reference image quality assessment (IQA) methods \cite{Kastryulin}. The no-reference IQA methods as well as full-reference IQA methods, appropriate for the assessment of CT projections and CT images, are yet to be developed.

Thus, the proposed self-supervised approach Noise2NoiseTD-ANM outperformed the other considered approaches in terms of adaptability and the quality of the denoised images. We believe this approach is promising given it can readily denoise LDCT images in a clinical setting. Further development of the self-supervised approach could decrease the dose of X-ray radiation even more, especially if combined with other low-photon image recovery methods \cite{pronina2020microscopy}. However, the proposed model still requires additional testing using more projections with real noise in a controlled manner. Lastly, the opinion of the radiologists about the projections denoised by our approach should also be analyzed.

\section*{Compliance with Ethical Standards}
This research study was conducted retrospectively using human subject data made available in open access by The Cancer Imaging Archive (TCIA). The usage of this dataset has been approved by the Philips internal committee for biomedical experiments.

\section*{Acknowledgments}
We thank Dr. Frank Bergner, Dr. Thomas Koehler (Philips GmbH Innovative Technologies) and Dr. Kevin M.Brown (Philips Healthcare) for supporting this research and providing feedback for this article.



\bibliographystyle{IEEEtran}
\bibliography{main.bib}

\begin{thebibliography}{10}
\providecommand{\url}[1]{#1}
\csname url@samestyle\endcsname
\providecommand{\newblock}{\relax}
\providecommand{\bibinfo}[2]{#2}
\providecommand{\BIBentrySTDinterwordspacing}{\spaceskip=0pt\relax}
\providecommand{\BIBentryALTinterwordstretchfactor}{4}
\providecommand{\BIBentryALTinterwordspacing}{\spaceskip=\fontdimen2\font plus
\BIBentryALTinterwordstretchfactor\fontdimen3\font minus
  \fontdimen4\font\relax}
\providecommand{\BIBforeignlanguage}[2]{{%
\expandafter\ifx\csname l@#1\endcsname\relax
\typeout{** WARNING: IEEEtran.bst: No hyphenation pattern has been}%
\typeout{** loaded for the language `#1'. Using the pattern for}%
\typeout{** the default language instead.}%
\else
\language=\csname l@#1\endcsname
\fi
#2}}
\providecommand{\BIBdecl}{\relax}
\BIBdecl

\bibitem{international19921990}
I.~C. on~Radiological~Protection, \emph{1990 Recommendations of the
  International Commission on Radiological Protection (Superseded by {ICRP}
  Publication 103): Adopted by the Commission in November 1990: User's
  Edition}.\hskip 1em plus 0.5em minus 0.4em\relax International Commission on
  Radiological Protection, 1992.

\bibitem{slovis2003children}
T.~Slovis, ``Children, computed tomography radiation dose, and the as low as
  reasonably achievable ({ALARA}) concept,'' \emph{Pediatrics}, vol. 112,
  no.~4, pp. 971--972, 2003.

\bibitem{brenner2007hall}
D.~J. D.J.~Brenner and E.~Hall, ``Computed tomography-—an increasing source
  of radiation exposure,'' \emph{New England Journal of Medicine}, vol. 357,
  no.~22, pp. 2277--2284, 2007.

\bibitem{SARMA2012750}
\BIBentryALTinterwordspacing
A.~Sarma, M.~E. Heilbrun, K.~E. Conner, S.~M. Stevens, S.~C. Woller, and C.~G.
  Elliott, ``Radiation and chest {CT} scan examinations: What do we know?''
  \emph{Chest}, vol. 142, no.~3, pp. 750--760, 2012. [Online]. Available:
  \url{https://www.sciencedirect.com/science/article/pii/S0012369212605212}
\BIBentrySTDinterwordspacing

\bibitem{yu2009radiation}
L.~Yu, X.~Liu, S.~Leng, J.~M. Kofler, J.~C. Ramirez-Giraldo, M.~Qu,
  J.~Christner, J.~G. Fletcher, and C.~H. McCollough, ``Radiation dose
  reduction in computed tomography: techniques and future perspective,''
  \emph{Imaging in medicine}, vol.~1, no.~1, p.~65, 2009.

\bibitem{3DCT}
H.~{Shan}, Y.~{Zhang}, Q.~{Yang}, U.~{Kruger}, M.~K. {Kalra}, L.~{Sun},
  W.~{Cong}, and G.~{Wang}, ``{3-D} convolutional encoder-decoder network for
  low-dose {CT} via transfer learning from a {2-D} trained network,''
  \emph{IEEE Transactions on Medical Imaging}, vol.~37, no.~6, pp. 1522--1534,
  2018.

\bibitem{RED}
H.~Chen, Y.~Zhang, M.~K. Kalra, F.~Lin, Y.~Chen, P.~Liao, J.~Zhou, and G.~Wang,
  ``Low-dose {CT} with a residual encoder-decoder convolutional neural
  network,'' \emph{{IEEE} Transactions on Medical Imaging}, vol.~36, no.~12,
  pp. 2524--2535, Dec. 2017.

\bibitem{ResCT}
W.~{Yang}, H.~{Zhang}, J.~{Yang}, J.~{Wu}, X.~{Yin}, Y.~{Chen}, H.~{Shu},
  L.~{Luo}, G.~{Coatrieux}, Z.~{Gui}, and Q.~{Feng}, ``Improving low-dose {CT}
  image using residual convolutional network,'' \emph{IEEE Access}, vol.~5, pp.
  24\,698--24\,705, 2017.

\bibitem{green2018learning}
M.~Green, E.~M. Marom, E.~Konen, N.~Kiryati, and A.~Mayer, ``Learning real
  noise for ultra-low dose lung {CT} denoising,'' in \emph{International
  Workshop on Patch-based Techniques in Medical Imaging}.\hskip 1em plus 0.5em
  minus 0.4em\relax Springer, 2018, pp. 3--11.

\bibitem{zabic2013low}
S.~{\v{Z}}abi{\'c}, Q.~Wang, T.~Morton, and K.~M. Brown, ``A low dose
  simulation tool for {CT} systems with energy integrating detectors,''
  \emph{Medical physics}, vol.~40, no.~3, p. 031102, 2013.

\bibitem{park2019unpaired}
H.~Park, J.~Baek, S.~You, J.~Choi, and J.~Seo, ``Unpaired image denoising using
  a generative adversarial network in {X}-ray {CT},'' \emph{IEEE Access},
  vol.~7, pp. 110\,414--110\,425, 2019.

\bibitem{wolterink2017generative}
J.~Wolterink, T.~Leiner, M.A.Viergever, and I.~Išgum, ``Generative adversarial
  networks for noise reduction in low-dose {CT},'' \emph{IEEE transactions on
  medical imaging}, vol.~36, no.~12, pp. 2536--2545, 2017.

\bibitem{kang2019cycle}
E.~Kang, H.~Koo, D.~Yang, S.~Seo, and J.~Ye, ``Cycle-consistent adversarial
  denoising network for multiphase coronary {CT} angiography,'' \emph{Medical
  physics}, vol.~46, no.~2, pp. 550--562, 2019.

\bibitem{n2n_ct2020}
P.~Gnudi, B.~Schweizer, M.~Kachelrieß, and Y.~Berker, ``Denoising of {X}-ray
  projections and computed tomography images using convolutional neural
  networks without clean data,'' in \emph{The 6th International Conference on
  Image Formation in X-Ray Computed Tomography}, 2020, pp. 590--593.

\bibitem{wu2019consensus}
D.~Wu, K.~Gong, K.~Kim, X.~Li, and Q.~Li, ``Consensus neural network for
  medical imaging denoising with only noisy training samples,'' in
  \emph{International Conference on Medical Image Computing and
  Computer-Assisted Intervention}.\hskip 1em plus 0.5em minus 0.4em\relax
  Springer, 2019, pp. 741--749.

\bibitem{hendriksen2020noise2inverse}
A.~A. Hendriksen, D.~M. Pelt, and K.~J. Batenburg, ``{N}oise2{I}nverse:
  Self-supervised deep convolutional denoising for tomography,'' \emph{IEEE
  Transactions on Computational Imaging}, vol.~6, pp. 1320--1335, 2020.

\bibitem{yuan2020half2half}
N.~Yuan, J.~Zhou, and J.~Qi, ``{H}alf2{H}alf: deep neural network based {CT}
  image denoising without independent reference data,'' \emph{Physics in
  Medicine \& Biology}, vol.~65, no.~21, p. 215020, 2020.

\bibitem{xu2021deformed2self}
J.~Xu and E.~Adalsteinsson, ``{D}eformed2{S}elf: Self-supervised denoising for
  dynamic medical imaging,'' in \emph{International Conference on Medical Image
  Computing and Computer-Assisted Intervention}.\hskip 1em plus 0.5em minus
  0.4em\relax Springer, 2021, pp. 25--35.

\bibitem{choi2021self}
K.~Choi, ``Self-supervised projection denoising for low-dose cone-beam {CT},''
  in \emph{2021 43rd Annual International Conference of the IEEE Engineering in
  Medicine \& Biology Society (EMBC)}.\hskip 1em plus 0.5em minus 0.4em\relax
  IEEE, 2021, pp. 3459--3462.

\bibitem{unal2021self}
M.~O. Unal, M.~Ertas, and I.~Yildirim, ``Self-supervised training for low-dose
  {CT} reconstruction,'' in \emph{2021 IEEE 18th International Symposium on
  Biomedical Imaging (ISBI)}.\hskip 1em plus 0.5em minus 0.4em\relax IEEE,
  2021, pp. 69--72.

\bibitem{zainulina2021n2ntd}
E.~Zainulina, A.~Chernyavskiy, and D.~V. Dylov, ``No-reference denoising of
  low-dose {CT} projections,'' \emph{arXiv preprint arXiv:2102.02662}, 2021.

\bibitem{laine2019hqss}
S.~Laine, T.~Karras, J.~Lehtinen, and T.~Aila, ``High-quality self-supervised
  deep image denoising,'' in \emph{Advances in Neural Information Processing
  Systems}, H.~Wallach, H.~Larochelle, A.~Beygelzimer, F.~d\textquotesingle
  Alch\'{e}-Buc, E.~Fox, and R.~Garnett, Eds., vol.~32.\hskip 1em plus 0.5em
  minus 0.4em\relax Curran Associates, Inc., 2019.

\bibitem{Noise2Noise}
J.~Lehtinen, J.~Munkberg, J.~Hasselgren, S.~Laine, T.~Karras, M.~Aittala, and
  T.~Aila, ``{N}oise2{N}oise: Learning image restoration without clean data,''
  in \emph{Proceedings of Machine Learning Research}, vol.~80.\hskip 1em plus
  0.5em minus 0.4em\relax PMLR, 2018, pp. 2965--2974.

\bibitem{krull2019noise2void}
A.~Krull, T.-O. Buchholz, and F.~Jug, ``{N}oise2{V}oid-learning denoising from
  single noisy images,'' in \emph{Proceedings of the IEEE/CVF Conference on
  Computer Vision and Pattern Recognition}, 2019, pp. 2129--2137.

\bibitem{xie2020noise2same}
Y.~Xie, Z.~Wang, and S.~Ji, ``{N}oise2{S}ame: Optimizing a self-supervised
  bound for image denoising,'' in \emph{Advances in Neural Information
  Processing Systems}, H.~Larochelle, M.~Ranzato, R.~Hadsell, M.~F. Balcan, and
  H.~Lin, Eds., vol.~33.\hskip 1em plus 0.5em minus 0.4em\relax Curran
  Associates, Inc., 2020, pp. 20\,320--20\,330.

\bibitem{batson2019noise2self}
J.~Batson and L.~Royer, ``{N}oise2{S}elf: Blind denoising by
  self-supervision,'' in \emph{International Conference on Machine
  Learning}.\hskip 1em plus 0.5em minus 0.4em\relax PMLR, 2019, pp. 524--533.

\bibitem{lee2020noise2kernel}
K.~Lee and W.-K. Jeong, ``{N}oise2{K}ernel: Adaptive self-supervised blind
  denoising using a dilated convolutional kernel architecture,'' \emph{arXiv
  preprint arXiv:2012.03623}, 2020.

\bibitem{buzug2011computed}
T.~M. Buzug, ``Computed tomography,'' in \emph{Springer handbook of medical
  technology}.\hskip 1em plus 0.5em minus 0.4em\relax Springer, 2011, pp.
  311--342.

\bibitem{la2006penalized}
P.~J. La~Rivi{\`e}re, J.~Bian, and P.~A. Vargas, ``Penalized-likelihood
  sinogram restoration for computed tomography,'' \emph{IEEE transactions on
  medical imaging}, vol.~25, no.~8, pp. 1022--1036, 2006.

\bibitem{liu2014dynamic}
F.~Liu, Q.~Yang, W.~Cong, and G.~Wang, ``Dynamic bowtie filter for
  cone-beam/multi-slice {CT},'' \emph{PloS one}, vol.~9, no.~7, p. e103054,
  2014.

\bibitem{shi2015convlstm}
X.~Shi, Z.~Chen, H.~Wang, D.~Y. Yeung, W.~K. Wong, and W.~C. Woo,
  ``Convolutional {LSTM} network: A machine learning approach for precipitation
  nowcasting,'' \emph{Advances in neural information processing systems}, vol.
  2015, pp. 802--810, 2015.

\bibitem{bias-free}
S.~Mohan, Z.~Kadkhodaie, E.~Simoncelli, and C.~Fernandez-Granda, ``Robust and
  interpretable blind image denoising via bias-free {C}onvolutional neural
  networks,'' in \emph{International Conference on Learning Representations},
  2020.

\bibitem{SE}
J.~Hu, L.~Shen, and G.~Sun, ``{S}queeze-and-{E}xcitation {N}etworks,'' in
  \emph{Proceedings of the IEEE Conference on Computer Vision and Pattern
  Recognition (CVPR)}, June 2018.

\bibitem{bromiley2003products}
P.~Bromiley, ``Products and convolutions of gaussian probability density
  functions,'' \emph{Tina-Vision Memo}, vol.~3, no.~4, p.~1, 2003.

\bibitem{zeng2015simulation}
D.~Zeng, J.~Huang, Z.~Bian, S.~Niu, H.~Zhang, Q.~Feng, Z.~Liang, and J.~Ma, ``A
  simple low-dose x-ray {CT} simulation from high-dose scan,'' \emph{IEEE
  transactions on nuclear science}, vol.~62, no.~5, pp. 2226--2233, 2015.

\bibitem{zhang2017dncnn}
K.~Zhang, W.~Zuo, Y.~Chen, D.~Meng, and L.~Zhang, ``Beyond a gaussian denoiser:
  Residual learning of deep {CNN} for image denoising,'' \emph{IEEE
  transactions on image processing}, vol.~26, no.~7, pp. 3142--3155, 2017.

\bibitem{ronneberger2015u}
O.~Ronneberger, P.~Fischer, and T.~Brox, ``U-net: Convolutional networks for
  biomedical image segmentation,'' in \emph{International Conference on Medical
  image computing and computer-assisted intervention}.\hskip 1em plus 0.5em
  minus 0.4em\relax Springer, 2015, pp. 234--241.

\bibitem{zhang2018ffdnet}
K.~Zhang, W.~Zuo, and L.~Zhang, ``{FFDN}et: Toward a fast and flexible solution
  for {CNN}-based image denoising,'' \emph{IEEE Transactions on Image
  Processing}, vol.~27, no.~9, pp. 4608--4622, 2018.

\bibitem{LDCT}
C.~McCollough, B.~Chen, D.~Holmes, X.~Duan, Z.~Yu, L.~Yu, S.~Leng, and
  J.~Fletcher, ``{L}ow {D}ose {CT} image and projection data
  ({LDCT}-and-{P}rojection-data),'' 2020.

\bibitem{TIGRE}
A.~Biguri, M.~Dosanjh, S.~Hancock, and M.~Soleimani, ``{TIGRE}: a
  {MATLAB}-{GPU} toolbox for {CBCT} image reconstruction,'' \emph{Biomedical
  Physics {\&} Engineering Express}, vol.~2, no.~5, p. 055010, sep 2016.

\bibitem{liu2014noise_level}
X.~Liu, M.~Tanaka, and M.~Okutomi, ``Practical signal-dependent noise parameter
  estimation from a single noisy image,'' \emph{IEEE Transactions on Image
  Processing}, vol.~23, no.~10, pp. 4361--4371, 2014.

\bibitem{xue2013gmsd}
W.~Xue, L.~Zhang, X.~Mou, and A.~C. Bovik, ``Gradient magnitude similarity
  deviation: {A} highly efficient perceptual image quality index,'' \emph{IEEE
  Transactions on Image Processing}, vol.~23, no.~2, pp. 684--695, 2013.

\bibitem{prokopenko2019unpaired}
D.~Prokopenko, J.~V. Stadelmann, H.~Schulz, S.~Renisch, and D.~V. Dylov,
  ``Unpaired synthetic image generation in radiology using {GAN}s,'' in
  \emph{Workshop on Artificial Intelligence in Radiation Therapy}.\hskip 1em
  plus 0.5em minus 0.4em\relax Springer, Cham, 2019, pp. 94--101.

\bibitem{Kastryulin}
S.~Kastryulin, J.~Zakirov, N.~Pezzotti, and D.~V. Dylov, ``Image quality
  assessment for magnetic resonance imaging,'' \emph{arXiv preprint
  arXiv:2203.07809}, 2022.

\bibitem{pronina2020microscopy}
V.~Pronina, F.~Kokkinos, D.~V. Dylov, and S.~Lefkimmiatis, ``Microscopy image
  restoration with deep wiener-kolmogorov filters,'' in \emph{European
  Conference on Computer Vision}.\hskip 1em plus 0.5em minus 0.4em\relax
  Springer, Cham, 2020, pp. 185--201.

\end{thebibliography}

\newpage

\section{Biography Section} 
 


\vspace{11pt}

\vspace{-33pt}
\begin{IEEEbiographynophoto}{Elvira Zainulina}
received the B.S. and M.S. degrees in applied mathematics and physics from Moscow Institute of Physics and Technology,
Moscow, Russia, in 2021, and the M.S. degree in mathematics and computer science from Skolkovo Institue of Science and Technology, Moscow, Russia, in 2019 and 2021, respectively. Starting from 2020 she joined Philips Research, Moscow, as a Junior Data Scientist. Her research interests include deep learning, computer vision, and medical image analysis.
\end{IEEEbiographynophoto}
\begin{IEEEbiographynophoto}{Alexey Chernyavskiy}
received his M.Sc. degree in applied mathematics and computer science from M.V. Lomonosov Moscow State University, Moscow, Russia, in 2000, a M.Sc. degree in geophysics from the University of Utah, Salt Lake City, USA, in 2003, and a Ph.D. degree in data analytics from the State Research Institute of Aviation Systems (GosNIIAS), Moscow, Russia, in 2013. Since 2019, he has been Senior Scientist at Philips Research, Moscow. His research interests include computer vision, image processing, and inverse problems.
\end{IEEEbiographynophoto}

\begin{IEEEbiographynophoto}{Dmitry V. Dylov}
received the M.Sc. degree in applied physics and mathematics from Moscow Institute of Physics and Technology, Moscow, Russia, in 2006, and the Ph.D. degree in Electrical Engineering from Princeton University, Princeton, NJ, USA, in 2010. 
He is an Associate Professor and the Head of Computational Imaging Group at Skoltech, Moscow, Russia. His group specializes on computational imaging, computer/medical vision, and fundamental aspects of image formation.
\end{IEEEbiographynophoto}

\vfill

\end{document}